\documentclass[twocolumn,pra,aps,superscriptaddress]{revtex4}

\usepackage{color}
\usepackage{amsmath}
\usepackage[dvips]{graphicx}
\usepackage{psfrag}
\usepackage{ulem}
\newcommand{\qed}{\hspace*{\fill}$\square$}

 \newtheorem{thm}{Theorem}
 \newtheorem{lem}[thm]{Lemma}
 
 \newtheorem{defn}[thm]{Definition}
 \newtheorem{prop}[thm]{Proposition}

 \newcommand{\C}{\mathbf{C}}
 
 \newcommand{\Z}{\mathbf{Z}}

 \newcommand{\funcion}[3]{#1:\,#2\longrightarrow #3}
 \newcommand{\img}{\mathrm {im}\,}

 \newcommand{\paratodo}{\forall\,}
 
 \newcommand{\vect}[1]{\boldsymbol{\mathrm{#1}}}
 \newcommand{\set}[2]{ \{\,#1\,|\,#2\,\}}
 \newcommand{\sset}[1]{ \{#1\} }
 \newcommand{\half}{\frac 1 2}

 \newcommand{\prima}{^\prime}
 \newcommand{\primas}{^{\prime\prime}}
 
 \newcommand{\nin}{ \not\in}

 \newcommand{\ket}[1]{|#1\rangle}
 \newcommand{\bra}[1]{\langle #1|}
 
 \newcommand{\ketbradif}[2]{\ket{#1}\bra{#2}}
 \newcommand{\ketbra}[1]{\ketbradif {#1}{#1}}

 \newcommand{\Hilb} {\mathcal H}

 \newcommand{\stable}{{\mathrm {dr}}}
 \newcommand{\syndrome}{\mathrm{synd}}
 \newcommand{\corrector}{\mathrm{corr}}
 \newcommand{\nodes}{\mathrm{nds}}
 \newcommand{\HilbEncoded}{\Hilb_{\mathrm{enc}}}
 \newcommand{\HilbExcitations}{\Hilb_{\mathrm{exc}}}

\def\ot{\otimes}
\def\hcal{{\cal H}}

\def\obs{A}
\def\virt{V}
\def\anc{Anc}
\def\psivecb{\psi_{\vect b}}

\newcommand{\<}{\langle}
\renewcommand{\>}{\rangle}
\renewcommand{\c}[1]{\mathcal{#1}}

\renewcommand{\r}[1]{\mathrm{#1}}
\renewcommand{\b}[1]{\mathbf{#1}}
\newcommand{\id}{\b 1}
\def\tr{\mathrm{Tr}}
\def\be{\begin{equation}}
\def\ee{\end{equation}}

\begin{document}

\title{ Self-Correcting Quantum Computers}

\author{H. Bombin}
\affiliation{Department of Physics, Massachusetts Institute of Technology, Cambridge, Massachusetts 02139, USA}
\author{R. W. Chhajlany}
\affiliation{Faculty of Physics, Adam Mickiewicz University, Pozna\'n{}, Poland}
\author{M. Horodecki}
\affiliation{Institute of Theoretical Physics and Astrophysics, University of Gda\'nsk, Poland}
\author{M.A. Martin-Delgado}
\affiliation{Departamento de F\'{\i}sica Te\'orica I, Universidad Complutense, 28040 Madrid, Spain}

\begin{abstract} Is the notion of a quantum computer resilient to thermal noise unphysical? We address this question from a constructive perspective and show that local quantum Hamiltonian models provide self-correcting quantum computers. To this end, we first give a sufficient condition on the connectedness of excitations for a stabilizer code model to be a self-correcting quantum memory. We then study the two main examples of topological stabilizer codes in arbitrary dimensions and establish their self-correcting capabilities. Also, we address the transversality properties of topological color codes, showing that 6D color codes provide a self-correcting model that allows the transversal and local implementation of a universal set of operations in seven spatial dimensions. Finally, we give a procedure to initialize such quantum memories at finite temperature.
\end{abstract}

\maketitle

\section{Introduction}

The quest for a robust and stable quantum computer (QC)
is a challenge in quantum information science and its
construction will imply the discovery of new physics.
We may classify QCs according to their protection
against different decoherence sources, as follows:
i/ Bare quantum computer. ii/ Externally protected QC.
iii/ Internally protected QC. This sequence ranges in
increasing degree of complexity. Case i/ corresponds to
an ideal functioning of a QC without errors. This was the
original formulation of the quantum circuit model prior 
to error correction \cite{nielsen_chuang}. The bare QC has already produced new
physics by means of small sized devices with ion traps, 
optical lattices and others \cite{cirac_zoller,nielsen_chuang}. Case ii/ is a big step forward
and corresponds to stabilizing the processing of quantum information
by means of acting externally on the QC in order to repair
the damage produced by errors and environment interactions.
Fault-tolerant quantum computing is the paradigm of this method
and the so called error-threshold theorem is its successful outcome,
meaning that quantum computation is not just analogue computation 
\cite{FTQC1,FTQC2,FTQC3, FTQC4a, FTQC4b, FTQC5, FTQC6, FTQC7}.
However, this approach has not met full experimental success yet.
Case iii/ is a more demanding approach in which we ask the 
quantum device to have the ability to correct itself whenever
an error occurs. Producing such a self-correcting QC will amount to finding 
a new quantum state of matter and topological orders \cite{wen_book} constitute promising 
candidates for this,  or variants thereof.

In this paper we take up the challenge of constructing 
a fully fledged self-correcting QC including
all basic operations: a) initialization, b) quantum gates, c) measurements.
We address the formulation of this problem from a fundamental viewpoint,
i.e., is  possible a proof-of-principle in order to satisfy
operations a), b) and c) in a self-correcting way?
We answer positively this question and give an explicit construction
of such an internally protected QC. Specifically, our solution is
constructed by means of a special type of topological quantum codes,
called topological color codes (TCC). To this aim, we construct
a new lattice with $D=6$ spatial dimensions and the desired properties in section \ref{sec:D-colexes}.
In order to appreciate the
implications of this result, 
we may draw a comprehensive parallelism between the current
achievements for externally and internally protected QCs.

This comparison 
states that, as long as the error rate $p$ of the externally
protected QC is low enough with respect to the threshold, then arbitrary long-term reliable
computations can be carried out. The bad news is that typical
error thresholds $p_c$ are very small in practice and no
quantum device of this type yet exists. Nevertheless, 
this result is considered a landmark of quantum computation,
a proof of principle that quantum processing is possible.
Interestingly enough, as for internally protected QCs, we have
a very similar situation, with pros and cons. We shall prove
that a complete self-correcting QC can be formulated as long
as the spatial dimension $D$ of the local array of qubits
is high enough. This is a remarkable result from a fundamental
point of view and sets the critical dimension of our
constructions to $D_c=6$. 
It opens the way to improve these constructions
in the direction of decreasing $D_c$, much like methods for
increasing $p_c$ are actively sought after.
An important remark is in order here: our models are soley constructed
upon qubits,  i.e., spin-$\half$ systems and symple interactions.
It is known though, that by changing the content and types of degrees of
freedom, a quantum system can lower its critical dimension. 
Thus, it is conceivable that more elaborate models than those
proposed here will be able to perform better as far as $D_c$ is
concerned. If this were not possible, then we would arrive at the
remarkable result that a self-correcting QC is not realizable in
a three-dimensional space lattice.

The issue of stability of self-correcting quantum memories \cite{Tcodes_1,Tcodes_2}
at finite temperature has been attracting much attention. One reason for this is that they are more physical
than the models of fault-tolerant quantum computation based
on the quantum circuit model \cite{AlickiLZ}. While the latter is based solely on
the naked quantum formalism with no dynamics, the topological memory relies
on a suitable Hamiltonian that protects quantum information as
a classical information is protected in, e.g., a  2D ferromagnet.
The original topological memories are placed on a torus, and it is
known that they offer stable qubits only starting from four spatial dimensions \cite{Tcodes_2,AlickiFH-Kitaev2,AlickiHHH-Kitaev,Kay-no-go,BravyiTerhal-no-go}. 

Both the toric models and others  \cite{Bacon06-3D} have some similar drawbacks. They do not support a set of universal gates, and
no scheme is known for the initialization of these self-correcting memories in the Hamiltonian spirit. Namely, one would like to
have something similar to an Ising model, where the state of definite
magnetization is prepared by simply switching on a global magnetic field
and then switching it off adiabatically.  Thus, disregarding for a while
the dimension of the model, an interesting question arises:
does it exist a self-correcting quantum computer, allowing
universal computation as well as a natural initialization scheme?

In this paper we propose a first such model. To this end we use color codes, a class of topological codes with very special transversality properties \cite{topologicalClifford,withoutBraiding}. Color codes are obtained from colexes, a class of lattices introduced in \cite{branyons}, and their underlying mathematical structure is a simplicial homology that differs from the usual one because it is `colored'. In the color code Hamiltonian models excitations take the form of branyons \cite{branyons}, punctual or extended objects subject to topological interactions. These models are local but require interactions between many spin-1/2 systems, a difficulty that can be overcome. E.g.,  Ref \cite{Zoller09-eff-local}
introduces an experimental proposal to implement the many-body interactions present both  in toric and color codes. Alternatively,
a local two-body Hamiltonian model that effectively yields a two dimensional color code has been also proposed \cite{twoBodyColor}.

The main idea of the paper is that in a suitable spatial dimension it is possible to find color codes that
support universal quantum computing and, at the same time, provide self-correction (which we prove giving a Peierls argument in the style of \cite{Tcodes_2,AlickiHHH-Kitaev}).
We also present a scheme of initialization, analogous to that of a ferromagnetic memory.
We believe that this is the first proposal of such an initialization,
as opposed to standard methods in fault tolerance (usually based on measuring
syndrome on some initial product state).  There is a remarkable difference
between our initialization scheme and that of a ferromagnetic memory.
Namely, we need to switch off at some point part of the protecting Hamiltonian,
which is not needed  in the latter case. Alternatively, we have proposed another
initialization method based in code deformation that does not require switching on/off the Hamiltonian (see Sect.\ref{sec:code_deformation}).

Let us emphasize that we are always using a realistic scenario where we always work at finite temperature $T>0$, implying that our 
notion of self-correctness  is really exact. We note that it is possible to come up with other notions of self-correction in which memories are initialized
at $T=0$ and demanding size-dependent thermalization time.

The paper is organized as follows. We first recall the  previous
results regarding thermal stability of self-correcting quantum memories (Sec. \ref{sec:results}).
 In Sec. \ref{sec:general} we discuss the subsystem picture, which is natural for
our analysis of thermal stability. In Sec. \ref{sec:stabilizer} we describe
general properties of stabilizer codes. We also introduce `dressed' observables -
the ones that define the protected qubit. In Sec. \ref{sec:noise} we present our modeling of the interaction with the environment. We recall an upper bound  on the decay rate of the autocorrelation functions, and its relation
with fidelity. We then formulate sufficient conditions for self-correction,
related to the Peierls argument in Sec. \ref{section_self_correction}. In addition, we prove that the given conditions also ensure that the quantum memory is resilient to a certain amount of noise applied suddenly.
Next we turn to the construction of the models, which must satisfy the conditions and at the same time allow a universal set of transversal operations
(Sec. \ref{sec:top-codes}). We give a unified homological picture for toric codes
and color codes, and prove in detail that some color codes possess
both properties at the same time. Finally, in Sec. \ref{sec:self-qc}
we propose a scheme for initialization, measurement and gate application.

\section{Previous results}
\label{sec:results}

The idea of a thermally stable self-correcting quantum memory
goes back to \cite{Tcodes_2}. The authors propose a 4D topological model and use a Peierls argument to provide heuristic basis in favour of its thermal stability. In \cite{AlickiHHH-Kitaev} this was proved rigorously using the formalism of quantum dynamical semigroups.
In this paper we generalize the latter proof. We give stability conditions that apply to general stabilizer codes, and then apply them to toric and color codes. 

Most likely there is no stable quantum memory in 2D. A full proof has been given in \cite{AlickiFH-Kitaev2} in the case of 2d Kitaev model (see also \cite{Kay2008-K2D} in  this context). The no-go results for stabilizer codes have been provided in \cite{Kay-no-go,BravyiTerhal-no-go}. Even though these results do not lead to a full proof of instability in the weak coupling limit, they point out quite clearly that there is no hope for stability in such models. 

Finally, let us note that many works have studied the existence of topological order in finite temperature (see e.g.  \cite{CastelnovoC08-order,IblisdirPAP08-order,Kargarian09-order}) and the main conclusions  were compatible with the results on stability of quantum memories.

\section{Subsystem approach to fault-tolerant quantum computing.}
\label{sec:general}

The most basic approach to quantum error correction is
the following. One considers a subspace, the code, chosen in such a way that, if the initial state
belongs to the subspace, and if not too many
errors hit the system, there exists a quantum operation,
the recovery map, that restores the initial state.

Such a scheme assumes ideal encoding and decoding, but in reality all operations are exposed to decoherence.
The prepared state is already erroneous to a certain extent.
Then, in fault-tolerant computing schemes,
it is subjected to repetitive error correction that keeps the amount of errors stable. At the end of computation, if the ideal recovery existed,
one could map the state back to the code subspace, but this is not really available nor needed. Rather, one wants to measure some observable
on the state. The measurement is nonideal, but if the state
did not suffer too many errors and the measurement does not produce too many either, then the
statistics will be arbitrarily close to those expected
in the ideal scenario, where encoding and recovery are perfect
operations.

To describe the above situation it is more convenient to consider a virtual subsystem \cite{KnillLV99-subsystems} (see also \cite{Alicki06-false-qubits} in this context). Then the fact that we cannot apply perfect
operations does not exclude that we can prepare the
subsystem in a given state with arbitrarily high fidelity.
For example, in standard hard drives, the sign of any single spin varies in time. However, the majority of signs of all spins  (i.e. magnetization) is
preserved with tremendous perfection.

The virtual subsystem is actually strictly related to the recovery procedure (see  \cite{Knill} in this context).  Namely,  the observables acting on the code
are lifted to the full space by subjecting them to an operation dual  to recovery map \cite{AlickiHHH-Kitaev} (we shall
present this construction in section \ref{sect_stable_observables}).

In the virtual subsystem picture, if we were able to start with the state within the code
the initial state would be $\psi_{\virt}\ot \phi_{\anc}^0$, where $\virt$ denotes the virtual subsystem and the second subsystem is $\hcal_{\anc}$, where $\psi$ is the wanted state of the subsystem,
then the errors (if not too many of them occur) affect only
the second subsystem, whose state turns into a mixed state.

In reality the encoding is not error-free,
so that the initial state is of the form
$|\psi\>\<\psi|_{\virt}\ot \rho^0_{\anc}$. If the state $\rho^0$ 
meets some conditions, then it is capable to accumulate the errors, which then do not affect 
the state $\psi_\virt$.  However, in absence of some additional mechanism, we will face some saturation,
and the errors will affect the systme $\virt$. 
In fault-tolerant schemes, there are two contradictory forces:
the environment adds errors (which propagate), while the user tries to correct errors. When these forces are balanced we get a dynamical equilibrium: the system is always
in a state of the form $|\psi\>\<\psi|_{\virt}\ot \rho_{\anc}$, i.e. the saturation 
will never occur. 

The above formulation is natural not only for
traditional fault-tolerant error correction schemes,
but also for self-correcting  models. In these models
there is a self-Hamiltonian that causes the
environment to do both jobs: adding and removing errors
(the latter due to Boltzmann factor). In such situation, the system
is (after initial equilibration) for a long time
in the state $|\psi\>\<\psi|_{\virt}\ot \rho_{\beta}$,
where $\rho_\beta$ is thermal equilibrium state
of the second subsystem.  The Gibbs state of the whole
system would be proportional to $I_{\virt} \ot \rho_\beta$,
and the  whole system will eventually relax to this state.
However, it may exist a critical temperature below which
the state of $\virt$ will be unaffected for a  time
that scales exponentially with the size of the total system \cite{Tcodes_2}.

In standard fault-tolerant schemes (where there is active error
correction), entropy is produced by
the environment and reduced by procedure of error correction.
In the self-correcting case both processes are performed by the environment. The fight (as usual
in phase transition phenomena) is between
entropy and energy - entropy causes errors and energy forces correction via Boltzmann factors.

Let us emphasize, that we do not deal here with the so called "noiseless
subsystem'' \cite{KnillLV99-subsystems}. The noiseless subsystem is protected against
some class of errors for {\it arbitrary} state of the rest of the system,
while here we have protection only when the rest is in
some particular state, e.g. Gibbs state, in the case of self-correcting
models. Also an error correcting code
usually does not form a noiseless subsystem:
assuming that the code can correct up to $k$ errors, only if we start
with initial pure state $\psi_{\virt} \ot \phi_{\anc}$ the subsystem  can indeed tolerate $k$ errors. Indeed to tolerate these errors we need to start from
the code, which in subsystem picture is the
span of $\psi_{\virt}\ot \phi_{\anc}$  for all $\psi$ from subsystem,
but for a fixed $\phi_{\anc}$.

The sketched subsystem picture is especially useful
if we want to analyse the interaction of the quantum memory with the environment. 
Then there are known tools to work with
autocorrelation functions of observables. To estimate stability
of the memory it suffices to show that these autocorrelations
do not decay in time. The suitable observables are
those from the algebra of the virtual subsystem, and their autocorrelation 
functions can be related to the fidelity of the state of the virtual subsystem. 

In section \ref{sec:stabilizer} we shall describe, first in the standard picture,
the error correction based on stabilizer formalism. Then we shall construct suitable 
`dressed'  observables which define the virtual subsystem.

\section{Stabilizer codes and transversal gates}
\label{sec:stabilizer}

\subsection{Stabilizer codes}

An error correcting code \cite{QEC_1,QEC_2} is a subspace $\mathcal
C$ of the Hilbert space $\Hilb$ that represents a quantum system.
The code subspace $\mathcal C$ is such that any quantum information
stored on it can be recovered after it has suffered errors from a
set $\mathcal E$ of correctable errors. Usually error correcting
codes are subpaces of systems of $n$ qubits, that is,
$\Hilb=\Hilb_2^{\otimes n}$ with $\Hilb_2$ a two-dimensional Hilbert
space with orthonormal basis $\sset{\ket 0, \ket 1}$. If $\mathcal
C$ encodes $k$ qubits, that is, has dimension $2^k$, and corrects
any error with support on less than $d/2$ qubits, we say that the
code is a $[[n,k,d]]$ code. The support of an operator is composed
of those qubits in which it acts nontrivially. In this context the
size of the support of an operator $O$ is usually called its weight
and denoted $|O|$.

An important class of error correcting codes is that of stabilizer
codes\cite{stabilizers_1,stabilizers_2}, which are defined in terms
of an abelian subgroup $\mathcal S\subset \mathcal P$ of the Pauli
group $\mathcal P$ of $n$ qubits. This is the group
\begin{equation}
\mathcal P=\langle iI, X_1, Z_1,\dots, X_n, Z_n\rangle,
\end{equation}
which has as generators the usual Pauli operators $X_i$, $Z_i$ on the $i$-th qubit,
\begin{equation}
X=\ketbra + - \ketbra -,\qquad Z=\ketbra 0 - \ketbra 1,
\end{equation}
where $\ket \pm:= (\ket 0 \pm \ket 1)/\sqrt 2$. The subgroups
$\mathcal S$ is called the stabilizer of the code $\mathcal C$
because its elements $\ket\psi\in \mathcal C$ are defined by
\begin{equation}
\paratodo s\in \mathcal S \qquad s\ket\psi =\ket\psi.
\end{equation}
Note that $\mathcal S$ cannot contain the element $-I$. The number of encoded qubits in a stabilizer code is
\begin{equation}\label{encoded_qubits}
k=n-g,
\end{equation}
with $g$ the number of independent generators of $\mathcal S$. To
give an expression for the distance of the code we have to introduce
the normalizer group $\mathcal N$. This is the normalizer of $\mathcal S$ in $\mathcal P$,
which coincides with the centralizer of $S$ in $\mathcal P$: its
elements $n\in \mathcal N\subset\mathcal P$ commute with all the stabilizers
$s\in\mathcal S$. Then the distance of the code $\mathcal C$ is
given by the minimum weight of the elements of $\mathcal N-\mathcal
S$, which indeed implement nontrivial unitary operations on encoded
states. In fact, it is possible to choose among them the $X$ and $Z$
Pauli operators for the $k$ encoded qubits: $\bar X_1, \bar Z_1,
\dots, \bar X_k, \bar Z_k\in \mathcal N-\mathcal S$. These operators
must satisfy the usual commutation relations, so that we have $\bar
X_i^2=\bar Z_i^2=1$, $\sset{\bar X_i,\bar Z_i}=0$ and $[\bar
X_i,\bar Z_j]=[\bar X_i,\bar X_j]=[\bar Z_i,\bar Z_j]=0$ for $i\neq
j$.

\subsection{Error correction}
If we put some quantum information into a given code subspace we
know that as long as the system does not suffer too many errors it
is in principle possible to recover the information without any
losses. But of course we are interested in recovering it in
practice, which is not necessarily just as easy.

Stabilizer codes have a particularly simple structure that makes
recovery from errors relatively easy. The first step in error
correction is the measurement of a set of generators $s_i$ of the
stabilizer
\begin{equation}
\mathcal S = <s_1,\dots, s_g>.
\label{eq:gen}
\end{equation}
Let us denote the eigenvalue of the $i$-th generator after the
measurement as $(-1)^{b_i}$, $b_i=0,1\in\Z_2$. The list of
eigenvalues $\vect b = (b_i)\in \Z_2^{g}$ is called the error
syndrome. The second step is then to choose, using the syndrome
$\vect s$, a Pauli operator $K\in\mathcal P$ such that
\begin{equation}\label{correction_opt_comm}
K s_i = (-1)^{b_i} s_i K.
\end{equation}
We set $\syndrome (K):=\vect b$ when \eqref{correction_opt_comm}
holds, so that $\syndrome$ defines a group homomorphism
\begin{align}
\funcion \syndrome {\mathcal P &} {\Z_2^g}\\
K &\longmapsto \vect b,
\end{align}
with kernel $\mathcal N$.

If the correction operator $K$ is applied to the system it will
return to the code subspace $\mathcal C$. Whether it returns to the
original state that we had encoded will depend on the errors that
occurred and the choice of $K$. To clarify this point, suppose for
simplicity that the action of errors amounts just to the application
of a Pauli operator $E\in\mathcal P$ to the system. Then we have by
assumption $KE\in \mathcal N$, and the error correction procedure
will succeed if in addition $KE\in \mathcal S$.

In summary, the whole error correction procedure is encoded in a
function
\begin{align}
\funcion \corrector {\Z_2^g &} {\mathcal P}\\
\vect b & \longmapsto K,
\end{align}
that takes syndromes $\vect b$ to correction operators $K$. This
$\corrector$ is such that
$$\syndrome(\corrector(\vect b))=\vect b.$$ One should choose
$\corrector$ in such a way that the probability of success for a
given source of errors is maximized. However, in practice it is also
important that $\corrector$ can be computed efficiently (say, in
polynomial time in the number $n$ of physical qubits), a requirement
that may reduce the success probability with respect to the ideal
case.

\subsection{Dressed observables and critical syndromes}\label{sect_stable_observables}

Take any nontrivial self-adjoint encoded operator $N\in \mathcal
N-\mathcal S$. Suppose that we initialize the system in an encoded
state and we want to keep track of the value of these encoded
operator as errors occur, from a theoretical perspective. In
particular, we would like to keep track of the value of $N$ were we
to correct the system using the above procedure. We may for this
purpose introduce the operator
\begin{equation}
\label{stable_opt}
N_\stable := \sum_{\vect b \in \Z_2^g} \corrector (\vect b)^\dagger
\,N\, \corrector (\vect b) \,P_{\vect b},
\end{equation}
where $P_{\vect b}$ is the projector onto the subspace with syndrome
$\vect b$, that is,
\begin{equation}\label{projector_syndrome}
P_{\vect b}:= 2^{-g}\prod_{i=1}^g \left (1+(-1)^{b_i} s_i \right ).
\end{equation}
Note that for encoded states $N_\stable=N$. The operator $N_\stable$
is a more stable version of $N$, because as long as the system only
suffers correctable errors its expectation value does not change.
This is not the case for $N$, which may change even with an error on
a single qubit.

Given an encoded operator $N$ it is useful to introduce the notion
of critical syndromes as those which are just a single-qubit error
apart from changing the value of $N_\stable$.
Let $\ket{\psivecb}$ denote a normalized state with $P_{\vect b}\ket{\psivecb}=\ket{\psivecb}$. 

For any Pauli operator $E\in\mathcal P$ we have
\begin{equation}
N_\stable \,E\, P_{\vect b} =S(N,E,\vect b) \,E N_\stable\, P_{\vect b},
\end{equation}
where
\begin{equation}
S(N,E,\vect b) =\pm 1
\end{equation}
is negative if and only if the encoded operator
\begin{equation}\label{trick_S}
\corrector (\vect b)\,E\,\corrector (\vect b+\syndrome (E))
\end{equation}
anticommutes with $N$.
Notice that for $E=E_1E_2$ we have
\begin{equation}\label{def_S}
S(N,E,\vect b)=S(N,E_1, \vect b+\syndrome (E_2))\,S(N,E_2, \vect b).
\end{equation}

When there exist a single-qubit Pauli operator $\sigma$ such that
$S(N,\sigma,\vect b)=-1$, we say that $\vect
b$ is $N$-critical. For each $N$, we denote by $\mathrm
{crit}_N\subset \Z_2^g$ the set of $N$-critical syndromes $\vect b$. Similarly, we denote by $\mathrm
{crit}\prima_N\subset \mathcal P\times\Z_2^g$ the set of pairs $(E,\vect b)$ with $S(N,E,\vect b)=-1$. Notice that for any $(E,\vect b)\in \mathrm{crit}_N\prima$ there exists a decomposition $E=E_1E_2$ such that $E_1$ and $E_2$ have disjoint supports and $\vect b+\syndrome (E_2)\in \mathrm
{crit}_N$.

Let us make connection with the general discussion from section \ref{sec:general}. 
The dressed observables $N_\stable$ define the virtual subsystem. 
They are built by means of observables that act solely on the code. Indeed,  
even though in expression \eqref{stable_opt} we use an observable $N$ that acts on the whole Hilbert space,
we use only its restriction to the code. Thus we obtained $N_\stable$ by lifting observables 
acting on the code  to those acting on the full system by using the correction procedure. 

The notion of critical syndrome refers to the 'capacity' of error storage 
on the ancilla system: when we are able to stay away from states with a critical syndrome, 
then our virtual system is protected. If the ancilla state a has critical syndrome, 
then one more added error affects the virtual subsystem. The system is self-correcting, 
if for a long time the ancilla system is far from a critical syndrome.

\subsection{Transversal gates}

An important feature of several classes of stabilizers codes is that
they allow the implementation of logical gates transversally, that
is, performing unitary gates on individual qubits or in small
disjoint sets of them. Usually we are interested in stabilizer codes
$\mathcal C$ that encode a single qubit. Then a one-qubit logical
gate $G$ that can be implemented in $\mathcal C$ through a unitary
operation of the form
\begin{equation}\label{transversal}
U=\bigotimes_i U_i
\end{equation}
is transversal. Here the index $i$ runs over physical qubits.
Frequently one has $U_i=U_j$, which means that no specific
addressing of the physical qubits is necessary, they are all treated
on equal footing.

A particularly simple case is that of $U$ belonging to the normalizer of $\mathcal P$, usually called the Clifford group. Then $U$ is a tranversal gate if for any $s\in \mathcal
S$
\begin{equation}\label{transversal_clifford}
U^\dagger s U\in \mathcal S,
\end{equation}
because for any $\ket \psi\in \mathcal C$ we must have $s U
\ket\psi=U\ket\psi$. An advantage of such tranversal gates is that, when applied with perfect accuracy, they affect the error syndrome of the code in a definite way. More generally it is enough for $U$ to satisfy
\begin{equation}\label{transversal_general}
U^\dagger P_{\vect 0} U= P_{\vect 0}.
\end{equation}
In section \ref{sec:transversal gates} we will consider such gates, which are essential for universal quantum computation. It is clear that from condition \eqref{transversal_general} we get no information about how $U$ affects the syndrome. However, \eqref{transversal} is a big constraint in this regard.
In particular, errors cannot spread. That is, if $\ket\psi$ is an encoded state and $E$ an error, then $U E \ket\psi = E\prima U \ket\psi$ for some $E\prima$ with the same support as $E$. A less trivial observation is that, in a certain sense, negative error syndromes also do not spread.  Suppose that we choose a possibly overcomplete set of stabilizer generators which are local in a given lattice of a certain dimension where the qubits live. Take a region $R$ of this lattice such that no encoded operator has a support completely contained on it. Then if $O$ is a Pauli operator that commutes with all the local generators that have part of their support in $R$, it follows as a slight generalization of the cleaning lemma of \cite{BravyiTerhal-no-go} that $O\ket\psi = O\prima\ket\psi$ for some $O\prima$ with support only outside $R$. This means that, if $E\ket\psi$ is free of negative error syndromes in a region, then the same will be true for $U E\ket\psi$ because $E\ket\psi = E\primas\ket\psi$ for some $E\primas$ with support outside $R$.

Similarly we can consider transversal two-qubit logical gates, which
are typically implemented on a pair of equal stabilizer codes
through unitaries of the form \eqref{transversal}, where now $i$
runs over pairs of equivalent physical qubits on the two codes, so
that the $U_i$ are two-qubit gates. Finally, it is also possible to
consider transversal measurements, where each physical qubit is
measured on a given basis and from the net result a measurement for
the encoded qubit is obtained.

\subsection{CSS-like codes}

CSS-like codes are those stabilizer codes for which the generators of the stabilizer can be chosen so that $\mathcal S=\langle s^X_1,\dots,s^X_{g_1},s^Z_1,\dots,s^Z_{g_2}\rangle$, $g_1+g_2=g$,
$$s_i^X\in \mathcal P_X, \qquad s_i^Z\in\mathcal P_Z,$$ with $\mathcal P_X := \langle
X_1,\dots,X_n\rangle$, $\mathcal P_Z := \langle
Z_1,\dots,Z_n\rangle$. In such codes, $X$-type ($Z$-type) logical operators may be chosen from $\mathcal P_X$ ($\mathcal P_Z$), that is, so that $$\bar X_i\in\mathcal P_X,\qquad \bar Z_i\in\mathcal P_Z.$$ Then $X$-type ($Z$-type) errors can only change the value of $Z$-type ($X$-type) encoded operators, and their correction only involves $Z$-type ($X$-type) stabilizers. That is, the error correction procedure can be divided into two subsystems, and the relevant functions are
\begin{align}\label{func_corr_synd_X_Z}
\funcion {\syndrome_Z} {\mathcal P_Z &} {\Z_2^{g_1}}, \funcion {\corrector_Z} {\Z_2^{g_1}} {\mathcal P_Z}, \nonumber\\
\funcion {\syndrome_X} {\mathcal P_X &} {\Z_2^{g_2}}, \funcion {\corrector_X} {\Z_2^{g_2}} {\mathcal P_X}.
\end{align}
Then for any $Z$-type encoded operator $N\subset \mathcal N\cap \mathcal P_Z$ the corresponding dressed operator takes the form
\begin{equation}\label{stable_opt_Z}
N_\stable := \sum_{\vect b \in \Z_2^{g_2}} \corrector_X (\vect b)^\dagger
\,N\, \corrector_X (\vect b) \,P^Z_{\vect b},
\end{equation}
where $P^Z_{\vect b}$ is the projector onto the subspace with $Z$-syndrome
$\vect b$, that is,
\begin{equation}\label{projector_syndrome_Z}
P^Z_{\vect b}:= 2^{-g_2}\prod_{i=1}^{g_2} \left (1+(-1)^{b_i} s^Z_i \right ).
\end{equation}
The dressed $X$-type Pauli observables are analogous.

An important consequence of the form of the dressed observables \eqref{stable_opt} is that transversal measurements are possible. Namely, if we measure all physical qubits in the $X$ basis ($Z$ basis) we can recover the value of all encoded $X$-type ($Z$-type) operators because error correction commutes with measurements. Indeed, after the measurement correction is entirely classical.

Finally, an analogous transversal initialization is possible. If we initialize a product state of the form $\ket 0^{\otimes n}$ ($\ket +^{\otimes n}$) and measure all $X$-type ($Z$-type) stabilizer operators, the resulting state has well defined $Z$-type ($X$-type) encoded operators. Namely, $\bar Z_i = 1$ (respectively, $\bar X_i = 1$). The state may have many $X$-type ($Z$-type) stabilizers with the wrong sign, but indeed the correction of $Z$-type ($X$-type) errors is immaterial because such errors cannot change the encoded state.
Of course this is the case  when  environment is absent. We shall discuss
initialization in the presence of noise in section \ref{subsec:init}.

\section{Modeling interaction with environment}
\label{sec:noise}

In this section we shall describe the evolution
of an open system in the weak coupling approximation.
Next we shall recall the relation between the fidelity criterion and the decay rate of observables,
and a useful upper bound for the latter \cite{AlickiHHH-Kitaev} (see also \cite{AlickiF-fidelity} in this context).

\subsection{Evolution in weak coupling limit}
We consider a quantum system with discrete energy spectrum is coupled to a collection of heat baths leading to the global Hamiltonian
\begin{equation}
 H = H^{\r{sys}} + H^{\r{bath}} + H^{\r{int}}
 \qquad\text{with}\qquad
 H^{\r{int}} = \sum_\alpha S_\alpha \otimes f_\alpha,
\end{equation}
where the $S_\alpha$ are system operators and the $f_\alpha$ bath
operators.

We assume that the baths are independent, which excludes such phenomena as decoherence free subspaces (see e.g. \cite{Alicki-DFS,ZanardiR-DFS,KnillLV99-subsystems}).
The evolution of the system in Heisenberg picture in weak coupling limit  is the following
\begin{align}
 \frac{dX}{dt}
 &= i[H^{\r{sys}},X] + \c L_{\r{dis}}(X) =: \c L(X)
\\
 \c L_{\r{dis}}(X)
 &= \frac{1}{2} \sum_\alpha \sum_\omega \hat h_\alpha(\omega) \Bigl(
 S_\alpha^\dagger(\omega)\, [X,S_\alpha(\omega)] +
 \\
 &+ [S_\alpha^\dagger(\omega),X]\,
 S_\alpha(\omega) \Bigr)
\label{gen}
\end{align}
Here $\hat h_\alpha$ are Fourier transforms of the autocorrelation functions of the $f_\alpha$. They describe the rate at which the coupling is able to transfer an energy $\omega$ from the bath to the  system.  In the case of thermal baths, they satisfy
\begin{equation}
 \hat h_\alpha(-\omega) = \r e^{-\beta\omega}\, \hat h_\alpha(\omega)
\end{equation}
which is a consequence of the KMS~condition \cite{Alicki-Lendi2}.
The dissipative part commutes with Hamiltonian part,
i.e. $[\hat H,\c L]=0$ (where $\hat H(X)=i[H^{sys},X]$).
Therefore in analysis of thermal stability,
it is enough to consider solely the dissipative part.
An important property of the dissipative generator is that
$-\c L$ is positive definite in scalar product
\begin{equation}
\<X,Y\>_\beta := \tr \rho_\beta\, X^\dagger\, Y.
\label{eq:product}
\end{equation}

Moreover we have $\c L(\id)=0$, and if  the commutant of the system operators  $S_\alpha$ and $H^{sys}$ is trivial,
then the eigenvalue 0 is nondegenerate. This means, that there are no constants of motion - all other observables eventually
decay to identity.

\subsection{Fidelity and autocorrelation functions}

Since the generator is hermitian in scalar product
\eqref{eq:product}, a strong tool is spectral analysis
of $\c L$. Namely, for any observable
satisfying $\<\obs,\obs\>_\beta=1$ and
$\<\obs, I\>_\beta=0$  one easily finds that
\be
\<\obs, \obs(t)\>_\beta \geq e^{-\epsilon t}
\ee
provided that
\be
-\< A, \c L (A)\>_\beta \leq \epsilon.
\ee
Thus, to show that an observable $\obs$ is stable,
it is enough to estimate $-\<\obs,\c L(\obs)\>$, which can be therefore called {\it decay rate} for the observable $\obs$.
If this quantity  decreases exponentially with size of the
system, we obtain stability.

However, the criterion of stability of quantum information
is measured in a most direct way by fidelity. Yet, in some
natural situations, the fidelity criterion
is equivalent to stability of two complementary observables
in terms of autocorrelation function. The needed relation was obtained in \cite{AlickiHHH-Kitaev}.
Let us recall it here, in a slightly more general setting. 

As discussed in section \ref{sec:general} we divide the total system 
into two: $\hcal_{\virt}\ot \hcal_{\anc}$,
where the first system is the qubit to be protected. 
We then consider an initial state of the form $|\psi_V\>\<\psi_V|\ot \rho_{anc}$.
The state is subjected to some completely positive trace preserving map $\Lambda^*_{V,anc}$,
and we are interested in the fidelity between the initial and final state of the system $V$:
\be
F(\psi_V;\Lambda^*_{V,anc},\rho_{anc})=
\<\psi_V|\rho_V^{out}|\psi_V\>
\ee
where $\rho_V^{out}=\tr_{anc}(\Lambda^*_{V,anc}(|\psi_V\>\<\psi_V|\ot \rho_{anc}))$.
In appendix \ref{appendix_fidelity} we prove the following proposition:
\begin{prop}
\label{prop:fid_virt}
Consider an arbitrary dichotomic observable $A_{V,anc}=A_V\ot I_{anc}$ acting on a Hilbert space 
$\hcal_V\ot \hcal_{anc}$ (where $\hcal_V=C^2$) with 
$\psi_\pm$ being eigenvectors of $A_V$ with eigenvalues $\pm1$. 
Then for an arbitrary state $\rho_{anc}$ and a completely positive trace preserving map $\Lambda^*_{V,anc}$
we have 
\be
F(\psi_+;\Lambda^*_{\virt,anc},\rho_{anc})+F(\psi_-;\Lambda^*_{\virt,anc},\rho_{anc})=1 + \<A,\Lambda(A)\>_\eta
\ee
where $\eta=\frac12 I_V\ot \rho_{anc}$, 
\be
\<A,B\>_\eta = \tr (\eta A^\dagger B),
\ee
and 
the map $\Lambda$ is dual to $\Lambda^*$.
\end{prop}
We shall also need the following lemma \cite{Hoffman}:
\begin{lem}
For an arbitrary trace preserving map $\Gamma$ on a qubit, 
define $F(\psi)= \<\psi|\Gamma(|\psi\>\<\psi|)|\psi\>$, as well as the 
entanglement fidelity (that quantifies the quality of preservation of a quantum memory)
\be
F_e(\Gamma)=\<\phi_+|(\Gamma \ot I)(|\phi_+\>\<\phi_+|)|\phi_+\>.
\ee
where $|\phi_+\>=\frac12(|00\>+|11\>)$. 
Then $F_e$ is bounded as follows :
\be
F_e(\Gamma)\geq F_{\sigma_x} + F_{\sigma_z} - 1
\ee
where $F_{\sigma_x}=\frac12 (F(|+\>) + F(|-\>))$, $F_{\sigma_z}=\frac12 (F(|0\>) + F(|1\>))$, with 
$|\pm\>$ being eigenvectors of $\sigma_x$, $|0\>,|1\>$ being eigenvectors of $\sigma_z$. 
\end{lem}
We shall now apply these facts to the present case.
Let $X$ and $Z$ be two Pauli observables
acting on the memory, i.e. they are of the form
$X_{\virt}\ot \id_{\anc}$ and $Z_{\virt}\ot \id_{\anc}$.
As the state $\eta$ in the proposition we take the Gibbs state 
(which is indeed proportional to
the identity on the system $\virt$ due to the form of Hamiltonian).
Finally the map $\Lambda$ is given by   $e^{\c L t}$. Applying the proposition and 
the lemma  we obtain 
\be
F_e\geq \frac12(\<X, e^{\c L t}X\>_\beta+\<Z, e^{\c L t}Z\>_\beta)\geq e^{-\epsilon t}
\ee
where $\epsilon$ is an upper bound for the rates $-\<X,\c L (X)\>_\beta$ and $-\<Z,\c L(Z)\>_\beta$.
Therefore, to prove stability of a quantum memory it is enough
to bound the above decay rates.
We have actually proved a more general result, namely that 
the  above inequality holds for whatever
physical operation (identity preserving completely positive map $\Lambda$) 
unrelated to the Hamiltonian of the system,
i.e. we have
\be
\label{eq:fidelity-autocor-general}
F_e \geq \frac12 (\< X, \Lambda(X)\>_\beta+\<Z, \Lambda(Z)\>_\beta).
\ee

\subsection{Upper bound for decay rate}

In \cite{AlickiHHH-Kitaev} the following useful bound for the
decay rate was obtained:
\be
-\<\obs,\c L(\obs)\>_\beta \leq 2 \hat h_{\max} \sum_\alpha
\<[S_\alpha, \obs],[S_\alpha, \obs]\>_\beta
\label{eq:gap}
\ee
where
\be
\hat h_{\max} = \sup_{\alpha, \omega\geq 0} \hat h_\alpha(\omega).
\ee
The advantage of this estimate is that it refers directly
to the system operators $S_\alpha$ rather to their Fourier transforms.

\section{Conditions for self-correcting quantum memories}
\label{section_self_correction}

\subsection{Stabilizer code models}

Given a stabilizer code, we may consider a Hamiltonian model of the
form
\begin{equation}\label{Hamiltonian}
H=-\sum_{i=1}^r t_i s_i\prima,
\end{equation}
where $t_i>0$ are coupling constants and $\sset{s_i\prima}$ is a possibly overcomplete set
of $r$ generators of the stabilizer $\mathcal S$. The encoded Pauli operators $\bar X_1, \bar Z_1,
\dots, \bar X_k, \bar Z_k$ commute with $H$, showing that each
energy level is at least $2^k$-fold degenerate. For a particular
choice of the encoded operators, one may split the Hilbert space
\begin{equation}\label{Hilbert_split}
\Hilb = \HilbEncoded\otimes \HilbExcitations = \C^{2k}\otimes
\C^{2g},
\end{equation}
in such a way that $\HilbEncoded$ represents the encoded degrees of
freedom and $\HilbExcitations$ characterizes excitations.
Let us emphasize that the above division is not yet the
one we considered in section \ref{sec:general}.
Indeed, the subsystem $\HilbEncoded$ is defined by the bare observables, and it is not protected.
In particular, we may choose a basis $\ket{\vect b}$ in
$\HilbExcitations$ with the elements labeled by $\vect b\in \Z_2^g$
so that setting $\ket{\psi, \vect b}:=\ket\psi \otimes \ket{\vect
b}$ for any $\ket\psi \in \HilbEncoded$ we have
\begin{equation}\label{basis_Ham}
s_i\,\ket{\psi, \vect b} = (-1)^{b_i} \ket{\psi, \vect b}.
\end{equation}
Recall that $s_i$ are generators from the complete set \eqref{eq:gen}.
Then the states of the form $\ket{\psi, \vect b}$ are eigenvectors
of the Hamiltonian \eqref{Hamiltonian}, namely
\begin{equation}\label{basis_Ham_energy}
H\, \ket{\psi, \vect b} = - \sum_{i=1}^r t_i^{c_i(\vect
b)}\ket{\psi, \vect b}=: E_{\vect b} \ket{\psi, \vect b},
\end{equation}
where $c_i(\vect b)=0,1$ are obtained from
\begin{equation}\label{stab_alternative}
s'_i = \prod_{j=1}^g s_j^{c_{ij}},\qquad c_{ij}=0,1,
\end{equation}
as $c_i(\vect b)=\sum_j c_{ij} b_j$, with addition modulo 2. Thus
the Gibbs state for the Hamiltonian \eqref{Hamiltonian} can be
written as
\begin{equation}\label{gibbs}
\rho(\beta) \propto \sum_{\vect b} e^{-\beta E_{\vect b}}P_{\vect
b}.
\end{equation}

\subsection{A bound in terms of critical syndromes}

In this section we shall show that the decay rate for a
dressed observable $N_\stable$ is bounded by the probability of a
critical syndrome multiplied by some polynomial in the size of the system. Thus, if the probability of the critical syndrome
is exponentially small in the size of the system,
the observable is stable.

Suppose that we have a Hamiltonian model \eqref{Hamiltonian} that
interacts with a bath through a coupling of the form
\begin{equation}\label{Hamiltonian_int}
H_{\mathrm {int}}=\sum_i X_i\otimes f_i + \sum_i Y_i\otimes g_i +
\sum_i Z_i\otimes h_i.
\end{equation}
For simplicity, we will write
\be
H_{\mathrm {int}}= \sum_\sigma
\sigma \otimes f_\sigma
\ee
 with $\sigma$ running over the three types
of Pauli operators and over all sites. We will suppose that our
model enjoys the following two properties. First, let the $t_i$
couplings satisfy 
\be
0<t_{\mathrm{min}}\leq t_i\leq t_{\mathrm{max}}.
\ee

Second, the number of stabilizers $s_i'$ (from \eqref{Hamiltonian})
that contain in their support a given qubit is bounded. When these
conditions hold, it follows that the Bohr frequencies $\omega$ 
are bounded, hance assuming that the spectral density $h(\omega)$ is finite for all $\omega$,
we obtain that  $\hat h_{\mathrm{max}}$ is finite and \eqref{eq:gap}
applies.

Let us go back to the dressed operators introduced in section
\ref{sect_stable_observables}. It is our aim to check how stable
they are under thermal fluctuations. From \eqref{eq:gap} and
\eqref{def_S} we have that for any encoded
operator $N$ the decay rate of $N_\stable$ is bounded as follows
\begin{multline}\label{decay_stable_op}
-\langle N_\stable, \mathcal L(N_\stable)\rangle_{\beta}\leq \\
\leq 8 \hat h_{\mathrm{max}} \frac {\sum_{\vect b} e^{-\beta
E_{\vect b}} \sum_{\sigma} (1-S(N,\sigma,\vect b))}{\sum_{\vect b}
e^{-\beta E_{\vect b}}}\leq \\ \leq 16 \hat h_{\mathrm{max}} n
P_{\mathrm {crit}_N}.
\end{multline}
Here $n$ is the total number of sites, which we will suppose that grows polynomially with the size of the system $L$, and $P_{\mathrm {crit}_N}$ is
the total probability at a given temperature of finding a syndrome
that is critical for $N$, that is
\begin{equation}\label{Probability_critical}
P_{\mathrm {crit}_N}=\frac  {\sum_{\vect b \in \mathrm
{crit}_N}e^{-\beta E_{\vect b}}} {\sum_{\vect b} e^{-\beta E_{\vect
b}}}.
\end{equation}
When for a given temperature $P_{\mathrm {crit}_N}$ is exponentially
small on the size of the system $L$ we say that $N_\stable$ is stable: it is self-protected by a free energy barrier
produced by the Hamiltonian \eqref{Hamiltonian}. In the following
section we will derive for a wide class of systems a set of
sufficient conditions for this stability to happen at some finite
temperature.

\subsection{Sketch of proof of stability of Kitaev model}
\label{sec:Kitaev3D-proof}

Before we present conditions assuring stability, let us first recall the essential feature
of the proof of stability of the Kitaev model \cite{AlickiHHH-Kitaev}. We shall restrict to a classical
submodel of the Kitaev model, where errors are only bit flips, and there is
therefore only one type of excitations. We consider configurations of
spins, i.e. patterns of spins-up or spins-down (or in bit notation,
this is a pattern of zeros and ones). These spins reside on the plaquettes of
some cubic lattice, while excitations reside on the edges. Any such
configuration leads to some pattern of excitations, which are closed
loops of edges. The energy associated with a given loop (excitation)
is proportional to its length.

The proof contains two main steps.

{\bf Step 1.} {\it Show that, with high probability only
  configurations leading to small loops appear in the Gibbs state.}

This is obtained as follows.  Consider a fixed loop, and assume that
it is homologically trivial.  Then by flipping some spins, we can
erase it from the configuration (without affecting any other loop). This means that we have a new spin
configuration, which differs from the old one {\it only} by the loop
that has been removed. The energy of the new configuration differs
from the old configuration only by the length of the erased loop.
This feature, i.e. possibility of {\it individually} erasing loops,
and the resulting energy difference between the respective spin
configurations, allows to prove, that the probability that a given loop
will appear, is bounded by $e^{-\beta l}$ where $l$ is the size of the
loop.  This is described in \cite{AlickiHHH-Kitaev}, 
following \cite{Griffiths1964-peierls}. Then the probability that arbitrary loops of length $l$ appear is bounded by
\be
p(l)\leq \#(loops) \times e^{-\beta l }.
\ee
where $\#(loops)$ is the number of all loops of length $l$.
So we have to count all  possible loops of length $l$.  We first bound the number of loops
containing a given spin.  Since loops are connected objects, it turns
out that this number is at most exponential in $l$, i.e. is $\leq
e^{\mu l}$ where $\mu$ is a constant dependent only on the type of
lattice involved.  So the entropic factor $\#(loops)$ is obtained by
multiplying this number by the volume $V$- {\it i.e.} the number of
spins in the lattice, which is polynomial in $L$ - the linear size of the
model.

Now this implies that the probability of appearance of loops of size
greater than  $c L$ (for some constant $c$) decays exponentially with
the system size $L$ for sufficiently low temperatures, as can be readily
seen by summing up the probabilities $p(l)$ from $l = cL$
upwards. Thus with high probability spin configurations will contain
only loops shorter than $c L$.

Apart from trivial loops typical spin configurations may also contain
non-trivial loops. These cannot be individually removed by flipping
spins, so the above argument and conclusion is not directly applicable
to these excitations (the dicussion of this case was actually missing in
\cite{AlickiHHH-Kitaev}). 
However the following device allows to
generalize the result for trivial loops to the case of non-trivial
loops. For the purpose of applying the {\it erasability} feature, note that
{\it groups} of non trivial loops can be erased by flipping spins. In
the case of the 3D Kitaev model, loops may be grouped (non-uniquely)
in to subsets of at most four loops 
(the number 4=3+1 comes from the $\Z_2$ first homology group of the 3D torus, 
which can be regarded as a vector space over $Z_2$ 
of dimension 3), such that each such subset can be erased as a whole by flipping spins.

Now, we can rerun the previous argument. Fixing some arbitrary {\it
erasable} set of no more than four loops, the probability of it
appearing is then again bounded by $e^{-\beta l}$ where $l$ now is the
size of the whole set.  The difference now is that the subset is not
connected which influences the counting of possible configurations
giving rise to a set of size $l$. However, a loose upper bound, that is
sufficient for our purposes, can be obtained by observing that the set
can have at most four components, each of which is a loop, i.e. a
connected object. Indeed, the number of such sets is trivially bounded from
above by the number of ways four loops of lengths
$l_{1},l_{2},l_{3},l_4$ can be chosen to form a subset of length
$l=l_{1}+l_{2}+l_{3}+l_4$, {\it i.e.}  $e^{\mu l} V^{4} poly(l)$
with $poly(l)$ the number of ways in which the individial lengths
 can be chosen to sum up to $l$.
Again summing probabilities, one obtains that
configurations that can be decomposed into erasable subsets of length
$l\geq cL$, for some constant $c$, appear with negligible probability for
sufficiently low temperatures.

{\bf Step 2.} {\it Show that configurations with short loops only
are not critical, i.e. single spin flip will not change the sign
of the dressed observable.}

This proof requires choosing a good error correcting algorithm 
(as the latter defines the dressed observable). Notice that, due to the previous discussion, only the correction of ``short" loops is relevant, e.g. of size smaller than, say, $L/2$. But all of these are homologically trivial and can be individually erased by flipping qubits living in the smallest paralelogram containing the relevant loop, which is univocally defined. The function $\corrector$ should be chosen accordingly.

Consider a configuration with short loops, such that each individual
loop has a size no greater than, say, $L/4$. A single flip affects only a few edges (4 in the
3-D Kitaev model) and thus can only affect a very small number of
loops (at most 2 in the 3-D Kitaev model) in a given
configuration. 
In the 3-D Kitaev model the flip can lead to the following changes:
(i)  'single loop $\leftrightarrow$ single loop' transition: alter the shape of a single loop along the flipped plaquette; 
(ii) 'two loops $\leftrightarrow$ one loop' transition: divide the loop into two
smaller loops or  join two loops into a single one; 
(iii) 'no loop $\leftrightarrow$ single loop' transitions: creation of a new single plaquette loop or anihilating such a existing small loop.
Thus if we take the product of the single quit bit-flip and the correcting operators before and after the change, the resulting product operator will have support on a cube of length $L/2+1$.
Hence it is a stabilizer and the configuration is not critical.

\subsection{A criterion based on connectedness}
\label{sec:conditions}

Here we present a set of conditions that assures that the probability
of a critical syndrome is exponentially small in the size of the
system. These conditions require the excitations to be connected in a
suitable way forming clusters, and also that critical configurations
always contain clusters of size comparable to the system size. This is
essentially an application of Peierls' argument
\cite{Peierls1936,Griffiths1964-peierls,Dobrushin1965-peierls} as
first done in \cite{Tcodes_2} in the context of quantum computation.
These conditions are obtained by extracting the essential
properties of the Kitaev 4D model (or the 3D
model for one kind of errors) used in the proof of its stability as recalled 
in the previous section.

For a given Hamiltonian $ H$, we shall consider an
error graph $\Gamma$, whose nodes represent the set $\sset
{s_i\prima}$ of generators of the stabilizer $\mathcal S$ appearing
in $ H $ \eqref{Hamiltonian}.  The nodes represent
excitations. Then any syndrome $\vect b\in\Z_2^g$ corresponds to a
subset $\nodes(\vect b)$. Namely, the $i$-th node belongs to the set
$\nodes(\vect b)$ if $c_i(\vect b) = 1$ (see eq. \eqref{stab_alternative}).
To be more explicit, an error causes a pattern of
excitations (it changes the sign of some generators from the
Hamiltonian) that depends on the
error only through the syndrome $\vect b$, so that we can unambiguously
write $\nodes(\vect b)$. While $\vect b$ tells us which independent
generators $s_i$ were affected by an error, $\nodes (\vect b)$ tells
us which Hamiltonian terms $s_i'$ were affected, which usually are not independent anymore.  The structure of links or edges in the graph will depend on the
particular model. It provides a notion of
connectedness: a set of nodes is connected if between any two of its
nodes there exists a path such that all its nodes belong to the
set. 
We say that $\vect b$ and $\vect b\prima$ contact if
$\nodes(\vect b)$ and $\nodes(\vect b\prima)$ are not disjoint or if
any of the elements of $\nodes (\vect b)$ is linked to an element of
$\nodes(\vect b\prima)$. 
Also,  given a Pauli operator $E$ we define $\nodes(E):= \nodes(\syndrome(E)).$

For example, in the 3D Kitaev model the qubits live in plaquettes and $X$ type excitations 
are associated with links. Such excitations are linked in the graph if the corresponding 
edges share a site of the original lattice. Connected sets take the form of closed paths.

Rather than a single Hamiltonian model \eqref{Hamiltonian}, consider a
family of models $H_L$ with increasing system size $L$, each with a
corresponding graph $\Gamma_L$. The number of nodes $|\Gamma_L|$ is
assumed polynomial in $L$.  We show that the following conditions, if
satisfied by the graphs $\Gamma_{L}$, imply the stability of the encoded
information. Here $\lambda , \mu , \nu $ are integers and $\xi $ a
positive number.

\noindent(i) For any single qubit Pauli $\sigma$ we have $|\nodes(\sigma)|\leq\lambda$.

\noindent (ii) The number of links meeting at a node is at most $\mu$.

\noindent (iii) For every $\vect b \in \Z_2^g$ there exists a decomposition
\begin{equation}\label{decomposition_syndrome}
\vect b = \sum_i \vect b_i
\end{equation}
such that the $\vect b_i$-s do not contact and
have at most $\nu$ connected components each.

\noindent (iv)
 If the total length of a syndrome satisfies $|\nodes (\vect b)|<\xi
 L$, for some number $\xi $, then the syndrome $\vect b$ is not
 critical for any encoded operator $N$, {\it i.e.}  $\vect b\nin
 \mathrm {crit}_N$.
Furthermore, for such $\vect b$, given any $\vect b\prima$ not contacting $\vect b$
\begin{equation}
\corrector (\vect b+\vect b\prima) = \corrector (\vect b)\,\corrector (\vect b\prima).
\end{equation}

Under these conditions one can characterize the type of
  syndromes that are critical w.r.t.  encoded information. Consider a
  general syndrome that can be decomposed into components
  (\ref{decomposition_syndrome}) of various lengths. We show that such
  a syndrome can be  critical only if it contains a large component $\vect
  b_0$ with $|\nodes (\vect b_0)|\geq(\xi L-\lambda)/\lambda\mu$.
Suppose, to the contrary that $\vect b = \sum_{i=1}^r \vect b_i$
consists of small components, {\it i.e.} 
  $|\nodes(\vect b_i)|<(\xi L-\lambda)/\lambda\mu$. 
  If $\sigma$ is a single-qubit Pauli operator, then
(i,ii) imply that at most $\lambda\mu$ of the $\vect b_i$-s make contact with $\nodes(\sigma)$, say $\vect b_1,\dots,\vect
b_k$. Let us denote $\tilde{\vect b} = \sum_{i=1}^k \vect b_i$.
Clearly we have $|\tilde{\vect b}|\leq \xi L-\lambda$ and  $|\tilde{\vect b}+ \syndrome(\sigma)|\leq \xi L$, hence
using (iv) we get
\begin{align}
\corrector (\vect b)&= \corrector(\tilde{\vect b})\prod_{i=k+1}^r
\corrector(\vect b_i),\nonumber\\ \corrector (\vect b+\syndrome
(\sigma))&= \corrector(\tilde{\vect b}+\syndrome(\sigma))\prod_{i=k+1}^r
\corrector(\vect b_i)
\end{align}
It follows that
\be
\corrector (\vect b) \sigma  \corrector (\vect b+\syndrome
(\sigma)) =
\corrector (\tilde {\vect b}) \sigma  \corrector (\tilde{\vect b}+\syndrome
(\sigma))
\ee
and hence due to  \eqref{trick_S} that $S(N,\sigma, \vect b)=S(N,\sigma,\tilde{\vect b})$,
i.e. $\vect b$ is critical  with respect to $\sigma$ if and only if $\tilde{\vect b}$
is critical  with respect to $\sigma$.
 However the latter satisfies $|\tilde{\vect b}|< \xi L$, hence
by (iv) it is not critical, so that  $\vect b \nin \mathrm {crit}_N$. In particular,
this result means that an error is not critical even if it
contains a large number of small error components. This result
corresponds to step 2 in the example of the 3-D Kitaev model above.

We now aim to bound the probability of a critical syndrome
  $P_{\mathrm {crit}_N}$ (much like in step 1 in the 3-D Kitaev
  model).  For this purpose, we first bound the probability $P_{\vect
    b_0}$ to find an excitation configuration $\nodes(\vect b)$ such
  that $\vect b_0$ is one of the components in the decomposition
  \eqref{decomposition_syndrome}. Denote the set of all configurations
  containing $\vect b_{0}$ by $\sset{\vect b_k\prima}$. We now apply
  Peierls' argument to show that the probability $P_{\vect b_{0}}$ is
  suppresed by the Boltzman factor at temperature $1/\beta$. Indeed,
  erasing the component $\vect b_{0}$ leads to configurations
  $\sset{\vect b_{k}\prima + \vect b_{0}}$ whose energies are related
  to the energies of the original configurations as $E_{\vect
    b_{k}\prima + \vect b_{0}}+E_{\vect b_{0}}=E_{\vect b_{k}\prima}
  $. Thus we obtain
\begin{multline}
P_{\vect b_0}=\frac {\sum_{k} e^{-\beta E_{\vect
b_k\prima}}}{\sum_{\vect b\in \Z_2^g} e^{-\beta E_{\vect b}}}=\frac
{e^{-\beta E_{\vect b_0}}\sum_{k} e^{-\beta E_{\vect b_k\prima+\vect
b_0}}}{\sum_{\vect b\in \Z_2^g} e^{-\beta E_{\vect b}}}\leq\\ \leq
e^{-\beta E_{\vect b_0}}\leq e^{-\beta t_{\mathrm{min}}
|\nodes(\vect b_0)|}.
\end{multline}
The fact that the energy corresponding to a syndrome depends only on
the lenght of the syndrome follows from the form of the Hamiltonian
(\ref{Hamiltonian}).
In order to obtain $P_{\mathrm {crit}_N}$, in accordance with the
previous paragraph, one needs the probability of appearance of any
error component of a large enough length.  The number $N(l)$ of
connected collections of nodes of size $l$ that contains a given node
is bounded as follows for some $\tau$ (see appendix \ref{appendix_connected}):
\begin{equation}\label{bound_N_l}
N(l)\leq e^{\tau l }.
\end{equation}
Thus, the total number of collections of $l$
nodes with at most $\nu$ connected components is bounded by
$|\Gamma_L|^\nu l^{\nu-1} e^{\tau l}\leq|\Gamma_L|^{2\nu}e^{\tau l}$.
Putting together these results we get
\begin{equation}\label{bound_P_crit}
P_{\mathrm {crit}_N}\leq |\Gamma_L|^{2\nu} \sum_{l\geq \xi
  L/\lambda\mu} e^{-\delta l}=\mathrm{poly}(L) e^{-\delta \xi
  L/\lambda\mu} \frac 1 {1-e^{-\delta}}
\end{equation}
with $\delta=\beta t_{\mathrm{min}}-\tau$ positive below a critical
temperature. This shows the stability of $N_{\stable}$ under the given
assumptions, as the critical probability decreases exponentially fast
with the system size.

\subsection{Resilience to noise}
\label{sec:noise_resilience}

\begin{figure}
  \centering
  \includegraphics[width=5cm]{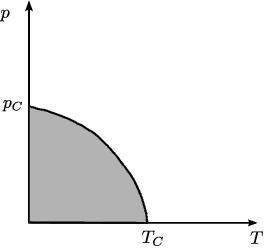}
  \caption{The shaded area describes schematically the protected region. $T_C$ is the critical temperature 
  in absence of noise, while $p_C$ is noise threshold for zero temperature.}
\label{fig:crit_line}
\end{figure}

It is natural to expect that a self-protecting quantum memory will
also be resilient, to some extent, to suddenly applied noise. This is
interesting since it allows us to expose the memory to errors, which will be an unavoidable byproduct of performing encoded gates. As we shall show, as long as the amount
of noise is below a threshold, we are safe. This threshold will reduce
as we increase the temperature till it vanishes at the critical
temperature, as schematically depicted on Fig. \ref{fig:crit_line}. 
For zero temperature, this is just the standard error threshold of 
the error correcting code.

To fix ideas, let us consider a process $\Lambda_p$ in which each qubit undergoes a depolarizing channel characterized by an `error' probability $p$. That is, any given operator $N$ is mapped in the Heisemberg picture to
\begin{equation}
\Lambda_p(N)= \sum_E (1-p)^{n-|E|}p^{|E|} \,E^\dagger N E,
\end{equation}
where the sum runs over a set of representants of $\mathcal P/iI$. We are under the error threshold when in the $L\rightarrow \infty$ limit for any given encoded operator $N$ we have
\begin{equation}
\langle N_\stable, \Lambda_p(N_\stable)\rangle_\beta \longrightarrow 1.
\end{equation}
But, setting $\eta:= \log((1-p)/p)$,
\begin{equation}
\langle N_\stable, \Lambda_p(N_\stable)\rangle_\beta = \frac {\sum_{E,\vect b} e^{-\beta E_{\vect b}-\eta |E|} S(N_\stable, E,\vect b)}{\sum_{E,\vect b} e^{-\beta E_{\vect b}-\eta |E|} }.
\end{equation}
That is, $\langle N_\stable, \Lambda_p(N_\stable)\rangle_\beta = 1-2 P_{\mathrm {crit}\prima_N}$ with
\begin{equation}
P_{\mathrm {crit}\prima_N}=\frac  {\sum_{(E,\vect b)\in \mathrm{crit}\prima_N} e^{-\beta E_{\vect b}-\eta |E|} }{\sum_{E,\vect b} e^{-\beta E_{\vect b}-\eta |E|} }.
\end{equation}
This expression is similar to \eqref{Probability_critical}, and we can use a similar reasoning to show that $P_{\mathrm {crit}\prima_N}\rightarrow 0$ for low enough temperature and noise.
First, we attach to each Pauli operator $E=\bigotimes_i \sigma_i$ the set $$\nodes'(E)=\bigcup_i \nodes(\sigma_i)\subset \Gamma_L$$ where the $\sigma_i$ are single qubit Pauli operators.
Notice that $|\nodes(E)|\leq|\nodes'(E)|\leq \lambda |E|$.
 Thus, when $\lambda|E|+|\nodes(\vect b)|<\xi L$ we have $(E,\vect b)\nin\mathrm{crit}\prima_N$.

Secondly, we need a substitute of the decomposition \eqref{decomposition_syndrome} to deal with pairs $(E,\vect b)$. 
To this end, notice that there exist $E_j$ such that $E=\prod_j E_j$ and the sets 
$\nodes' (E_j)$ form the connected components of $\nodes'(E)$.
 Then we arrange the $\vect b_i$-s and $E_j$-s in maximal connected collections. That is, if one of these collections is $\sset {\vect b_{a_1},\dots \vect b_{a_r}, E_{a_1}, \dots, E_{a_s}}$, then the set $\nodes(\vect b_{a_1})\cup \dots\cup \nodes(\vect b_{a_r})\cup\nodes'(E_{a_1})\cup \dots\cup \nodes'(E_{a_r})$ is connected and disconnected from any set $\nodes (\vect b_i)$ or $\nodes'(E_j)$ that is not part of the collection. 
 We denote each collection as $(E_a, \vect b_a)$, with $\vect b_a=\sum_i\vect b_{a_i}$, $E_a=\sum_i E_{a_i}$, so that $\vect b=\sum_a\vect b_a$ and $E=\prod_a E_a$. This gives the desired decomposition of $(E,\vect b)$.

Finally, we observe that when $\lambda|E_a|+|\nodes(\vect b_a)|<\xi L$ for all pairs $(E_a,\vect b_a)$ we have
\begin{align}
\corrector (\vect b)&= \prod_a \corrector(\vect b_a),\nonumber\\
\corrector (\vect b+\syndrome (E))&= \prod_a\corrector(\vect b_a+\syndrome(E_a)),
\end{align}
so that $S(N,E, \vect b)=\prod_a S(N,E_a, \vect b_a)$. We omit the rest of details.

\section{Topological stabilizer codes}
\label{sec:top-codes}

In this section we shall construct codes that  allow a universal set of transversal gates and, at the same time, give rise to self-correcting local Hamiltonian models. The codes will belong to the important class of local stabilizer codes, in particular topological codes. These are constructed from manifolds where the physical qubits are placed, and have two main characteristics:
\begin{itemize}
 \item
 There exists a set of generators of the stabilizer $\mathcal S$ with a spatially local support.
 \item
 The encoded operators in $\mathcal N-\mathcal S$ are global, they have a support which is topologically nontrivial.
\end{itemize}
We will give below a more precise characterization. The quantum
Hamiltonian models \eqref{Hamiltonian} obtained from topological
codes are topologically ordered: their ground spaces have a
degeneracy of topological origin and their excitations are gapped
and subject only to topological interactions. As we will see,
excitations can be either particles or higher dimensional objects.
Self-correction will appear in codes with excitations of the latter kind.

The original example of topological stabilizer codes are toric
codes, as introduced by Kitaev\cite{Tcodes_1,Tcodes_2}. These codes
and their generalizations are based on homology
theory\cite{homological}. Indeed, the number of encoded qubits in a
given manifold is dictated by its Betti numbers, i.e., the
dimensions of homology groups, which are
topological invariants.

An important drawback of toric codes is that they do not allow the
transversal implementation of a sufficiently rich set of gates.
There exists however a class of topological stabilizer codes that
surmount this difficulty: color codes. In 2D
\cite{topologicalClifford} they allow the transversal implementation
of the so-called Clifford group, the normalizer of the Pauli group
in the group of unitary operations on $n$ qubits. In 3D
\cite{withoutBraiding} they allow  the transversal implementation of
a set of operations that, when measurements are included, gives rise
to transversal universal quantum computation. Below we will analyze
in detail the transversality properties of general $D$-dimensional
color codes.

\subsection{Local stabilizer codes}

A family of codes $\sset{\mathcal C_L}$ is local if there exists
integers $\alpha$ and $\beta$ such that

\noindent (i) each stabilizer $\mathcal S_L$ has a set of generators
$\sset{s_1,\dots,s_{g_L}}$ such that $|s_j|\leq \alpha$, $j=1,\dots,
g_L$,

\noindent (ii) each physical qubit is in the support of at most
$\beta$ such generators $s_j$ and

\noindent (iii) for any integer $d$, there exist a code $\mathcal
C_L$ with distance bigger than $d$.

\noindent Given a code $\mathcal C_L$ in the family, it is always
possible to construct a local quantum Hamiltonian model such that
its ground state is $\mathcal C_L$. Namely, we may set
\begin{equation}\label{Hamiltonian_local}
H_L=-\sum_{i=1}^{g_L} s_i.
\end{equation}

Topological stabilizer codes are local codes in which locality has a geometrical meaning. Physical qubits are placed in a lattice in a
given manifold, in such a way that the generators of the stabilizer
have a support contained in a small neighborhood of the lattice. The
support of nontrivial encoded operators, in turn, must be global, in
such a way that the distance of the codes grows with the size of the
lattice \cite{compareTopoCodes}.

Below we will study in full generality two important examples of
topological stabilizer codes, toric codes and color codes. We will
unify the treatment of both types of codes through homology theory.

\subsection{Homology groups}

Recall that a homology group is defined from an exact sequence such
as
\begin{equation}\label{exact_sequence}
A_2 \,\xrightarrow {\partial_2}\,A_1\,\xrightarrow {\partial_1}\,A_0
\end{equation}
where $A_i$ are Abelian groups and $\partial_i$, the boundary
operators, are homomorphisms with
\begin{equation}\label{closed_if_exact}
\partial_1 \circ \partial_2 = 0.
\end{equation}
The elements of the subgroup $Z_1:=\ker \partial_1\subset A_1$ are
called cycles. The elements of the subgroup $B_1:=\img
\partial_2\subset A_1$ are called boundaries. Then the condition
\eqref{closed_if_exact} states that all boundaries are cycles,
$B_1\subset Z_1$. Because of this, we can divide cycles in homology
classes. Two cycles $z$, $z\prima$ are homologous, $z\sim z\prima$,
if $z-z\prima$ is a boundary. The homology group $H$ has as its
elements the corresponding homology classes. It is the quotient
\begin{equation}\label{defn_homology_group}
H := {Z_1}/{B_1}.
\end{equation}

In our case, the elements of the groups $A_i$ will be chains of suitable geometrical
objects from a lattice in a given manifold, and the homology group will be a topological invariant of the manifold, independent of the particular lattice considered. We are only interested in $\Z_2$ homology, in which the
chains in $A_i$ are binary formal sums of geometrical objects in a
set $S_i$. That is, if $a\in A_i$ then
\begin{equation}
a=\sum_{s\in S_i} b_s \,s, \qquad b_s=0,1.
\end{equation}
When needed, we will regard $\Z_2$ chains as sets in the usual way.
That is, we identify the chain $a$ above with the set $\set{s\in
S_i}{b_s=1}$.

When working with such $\Z_2$ homology groups we may consider the dual exact sequence
\begin{equation}\label{exact_sequence_dual}
A_0 \,\xrightarrow {\partial_1^\ast}\,A_1\,\xrightarrow {\partial_2^\ast}\,A_2.
\end{equation}
The dual boundary operators $\partial_i^\ast$ are defined so that for $a\in S_{i-1}$, $b\in S_i$ we have $a\in\partial_i b$ if and only if $b\in\partial_i^\ast a$. The dual exact sequence \eqref{exact_sequence_dual} gives rise to a dual homology group $H^\ast:=Z_1^\ast/B_1^\ast$ which is isomorphic to $H$. Notice that for $a\in A_1$, $s_2\in S_2$ and $s_0\in S_0$  we have
\begin{align}\label{parity_vs_boundary}
s_0\in\partial_1 a \iff |\partial_1^\ast s_0 \cap a|\equiv 1 \mod 2,\nonumber\\
s_2\in\partial_2^\ast a \iff |\partial_2 s_2 \cap a|\equiv 1 \mod 2.
\end{align}

\subsection{Codes from homology groups}\label{section_homology_codes}

Given a $\Z_2$ homology group $H$ obtained from a lattice in a manifold, as those discussed in the previous section, we can introduce in a natural way a CSS-like stabilizer code $\mathcal C$ with a basis labeled by the elements of $H$. The procedure can be summarized as follows. Recall that the abelian groups $A_i$ in the exact sequence \eqref{exact_sequence} are constructed from sets of geometrical objects $S_i$. To define the code, we attach physical qubits to the elements of $S_1$, $X$-type stabilizers to the elements of $S_2$ and $Z$-type stabilizers to the elements of $S_0$. In what follows, we detail the procedure.

First, we attach a physical qubit to each of the elements of $S_1=\sset{r_j}$, so that the number of qubits $n$ is given by $n=|S_1|$.
For each $a\in A_1$ with
\begin{equation}
a=\sum_{j} a_j \, r_j, \qquad a_j = 0,1
\end{equation}
we define a $X$-type and a $Z$-type Pauli operator setting
\begin{equation}\label{chain_opt}
X_a := \bigotimes_j X_j^{c_j},\quad Z_a := \bigotimes_j Z_j^{c_j}, \qquad a =\sum_j c_j r_j.
\end{equation}
Note that the definitions \eqref{chain_opt} describe two group isomorphisms. Moreover,
\begin{equation}\label{chain_opt_comm}
X_a Z_{a\prima} = (-1)^{|a\cap a\prima|} Z_{a\prima} X_a.
\end{equation}
Then using \eqref{parity_vs_boundary}
\begin{align}\label{exact_closed_comm}
a\in Z_1 \iff \forall b\in B_1^\ast,\quad  &[X_a, Z_b]=0 \\ a\in Z_1^\ast \iff \forall b\in B_1, \quad &[Z_a, X_b]=0.
\end{align}
As a consequence of \eqref{closed_if_exact}, $[X_b,Z_{b\prima}]=0$ for $b\in B_1,b'\in B_1^*$ and we can define the stabilizer group $\mathcal S$ as the subgroup of $\mathcal P$ with elements of the form
\be X_bZ_{b\prima},\qquad b\in B_1, b\prima\in B_1^\ast.
\ee
It follows then that the operators of the form \be X_zZ_{z\prima},\qquad z\in Z_1, z\prima\in Z_1^\ast,\ee form, up to  phases $\pm1,\pm i$, the normalizer group $\mathcal N$. If $\bar X_1,\dots,\bar X_k\in\mathcal P_X$ and $\bar Z_1, \dots, \bar Z_k\in\mathcal P_Z$ form a Pauli basis for encoded operators, then $X_i= X_{z_i}$ and $Z_i=Z_{z\prima_i}$ for suitable cycles $z_i\in Z_1$, $z_i\prima\in Z_1^\ast$. The cycles $z_i$ ($z_i\prima$) are representants of classes $\bar z_i\in H$ ($\bar z_i\prima\in H^\ast$) that form an independent set of generators of $H$ ($H^\ast$), and thus we have the desired result.

It is instructive to work out explicitly a basis for the resulting code $\mathcal C$. To this end, for any $a\in A_1$ we set
\begin{equation}\label{basis_0_plus}
\ket a_0 :=X_a \,\ket 0^{\otimes n}, \qquad \ket a_+ :=Z_a \,\ket +^{\otimes n}.
\end{equation}
Then $\sset {\ket a_0}$ and $\sset {\ket a_+}$ are bases of the Hilbert space, with elements labeled by the chains $a\in A_1$. For any cycles $z\in Z_1$, $z\prima\in Z_1^\ast$ we define the states
\begin{equation}\label{code_states}
\ket {\bar z}_0 := \sum_{c \sim z} \ket {c}_0, \qquad \ket {\bar z\prima}_+ := \sum_{c \sim z\prima} \ket {c}_+.
\end{equation}
Then, as can be easily checked, $\sset {\ket {\bar z}_0}$ and $\sset {\ket {\bar z\prima}_+}$ are alternative bases for the topological stabilizer code $\mathcal C$ with elements labeled by homology classes of $H$ and $H^\ast$, respectively.

\subsubsection{Error correction}

We analyze now error correction in the code $\mathcal C$ just described. First, we need a generating set for the stabilizer group, which can be obtained by attaching an operator to each of the elements of $S_2$ and $S_0$. Set for any $a_2\in A_2$, $a_0\in A_0$,
\begin{equation}
X_{a_2} := X_{\partial_2 a_2}, \qquad Z_{a_0} := Z_{\partial_1^\ast a_0}.
\end{equation}
Then if $S_2=\sset{p_1,\dots,p_l}$, $S_0=\sset{q_1,\dots, q_m}$, we may write
\begin{equation}\label{generators_X_Z}
\mathcal S := \langle X_{p_1},\dots,X_{p_l},Z_{q_1},\dots,Z_{q_m}\rangle.
\end{equation}
Indeed, the above set may be overcomplete. We can choose in that case suitable subsets $S_2\prima=\sset{p_1,\dots,p_{g_1}}\subset S_2$, $S_0\prima=\sset{q_1,\dots,q_{g_2}}\subset S_0$. We may naturally regard any $s_2\subset S_2\prima$ as some $\vect b\in\Z_2^{g_1}$, setting $b_i=1$ when $p_i\in s_2$, and similarly for $S_0\prima$ and $\Z_2^{g_2}$. Then, for $a\in A_1$,
\begin{align}\label{syndrome_topo_codes}
\syndrome_X(Z_{a})&= S_2\prima\cap \partial_2^\ast a,\nonumber \\
\syndrome_Z(X_a)&= S_0\prima\cap \partial_1 a.
\end{align}
That is, for a given Pauli error $X_aZ_{a\prima}$ the error syndrome determines the boundaries of $a$ and $a\prima$. Error correction amounts to choose an operator $X_{c}Z_{c\prima}=\corrector(\syndrome(X_aZ_{a\prima}))$ with $\partial_1 c=\partial_1 a$ and $\partial_2^\ast c\prima=\partial_2^\ast a\prima$. It will succeed if and only if $a+c\in B_1$ and $a\prima+c\prima\in B_1^\ast$, that is, if and only if the correction operator and the error are equivalent up to homology.

\subsubsection{Hamiltonian and excitations}

Given the generators of the stabilizer in \eqref{generators_X_Z}, we can consider the quantum Hamiltonian model
\begin{equation}\label{Hamiltonian_X_Z}
H=-\sum_{i=1}^{l} X_{p_i}-\sum_{i=1}^{m} Z_{q_i}.
\end{equation}
The excitations of this model are gapped, with two units of energy per stabilizer violation. We say that the excitations related to $X_{p_i}$ ($Z_{q_i}$) generators are $X$-type ($Z$-type) excitations. Then, as follows from the previous analysis of error correction, $X$-type excitations are arranged in the form of $B_1^\ast$ boundaries and $Z$-type excitations are arranged in the form of $B_1$ boundaries.

\subsection{Generalized toric codes}

Consider a lattice in a $D$-manifold. We call the $n$-dimensional
elements of the lattice $n$-cells: $0$-cells are vertices, $1$-cells
are links, $2$-cells are faces, and so on. For $n=0,\dots, D$, let
$S_n$ be the set of all $n$-cells, and $C_n$ the corresponding
abelian group of chains. We in addition set $C_{D+1}=C_{-1}=0$, with
$0$ the trivial group. For $n=1, \dots D$ we introduce the boundary
operators
\begin{equation}
C_n \,\xrightarrow {\partial_{n}}\,C_{n-1}
\end{equation}
in the usual way. That is, they are homomorphisms such that if the
$(n-1)$-cells $s_i\in S_{n-1}$ form the geometrical boundary of the
$n$-cell $s$ then $\partial s =
\sset{s_i}$. We also define the homomorphisms $\partial_{D+1}$ and
$\partial_{0}$ in the only possible way.

We can obtain a generalized toric code for each $d$ with $d=0,\dots, D$. This is the code that corresponds to the exact sequence
\begin{equation}\label{seq_toric}
C_{d+1} \,\xrightarrow {\partial_{d+1}}\,C_{d}\,\xrightarrow {\partial_{d}}\,C_{d-1},
\end{equation}
as described in the previous section. The corresponding homology group is the standard $d$-th homology group of the manifold,
\begin{equation}\label{homology_d}
H_d := \frac {Z_d}{B_d}.
\end{equation}
Thus the number of encoded qubits $k$ is $h_d$, the $d$-th Betti number of the manifold. For example, for $d=1$ we have the homology group of curves, so that the number of encoded qubits is equal to the number of independent non-boundary loops in the manifold. Notice that the $0$-th and the $D$-th code are just classical repetition codes, with one encoded qubit per connected component of the manifold.

The dual sequence can be visualized more easily in the dual lattice, in which $r$-cells become $(D-r)$-cells. Indeed, in the dual lattice the dual exact sequence corresponds to the exact sequence of the $\bar d$-th homology group, where
\begin{equation}
 \bar d:=D-d.
\end{equation}
Thus, in a $d$-th toric code the relevant homology groups are $H_{\bar d}$ and $H_d$. Cycles take the form thus of closed $\bar d$- and $d$-branes, which are mapped to operators using \eqref{chain_opt}. The commutation rules of closed brane operators are purely topological. Namely, given a closed $d$-brane $z\in Z_d$ and a closed
$\bar d$-brane $z\prima\in Z_{\bar d}$, the operators $X_z$ and $Z_{z\prima}$ anticommute if $z$ and $z\prima$ intersect and odd number of times. Otherwise they commute.

\subsubsection{Branyons}

For $d=1,\dots,D-1$, the Hamiltonian model \eqref{Hamiltonian_X_Z} of a $d$-th toric code contains two kinds of brane-like excitations, which are created by open $d$ and $\bar d$-brane operators. $Z$-type excitations are closed $(d-1)$-branes and $X$-type excitations are closed $(\bar d-1)$-branes in the dual lattice. Indeed, both kind of excitations are not only closed but also boundaries. For example, for $D=2$ and $d=1$ the excitations are $0$-branes, that is, particles. They can only be created or destroyed in pairs since they have to be boundaries of strings. These two kinds of particles interact topologically and are thus called anyons. More generally, $Z$-type $(d-1)$-branes and $X$-type $(\bar d-1)$-branes interact topologically and thus deserve the name of branyons \cite{branyons}.

Since $(n-1)$-branyons appear at the boundaries of $n$-brane operators, $n=d,D-d$, it follows that such operators represent branyon processes. In particular, processes that do not create or destroy excitations are related to closed branes. Then one can derive from the commutation rules for closed brane operators the topological $\pm 1$ phases appearing in branyon processes.

\subsubsection{Self-correcting models}
\label{subsubsec-self-toric}

To fix ideas, let us consider a family of hypercubic latices in a $D$-dimensional torus. Each lattice has the same lenght $L$ in the $D$ directions. We will show that, for $d> 1$, the models with Hamiltonian \eqref{Hamiltonian_X_Z} for the $d$-th code protect $Z$-type encoded operators in the sense of section \ref{section_self_correction}. Then it follows by duality that for $d< D-1$ $X$-type operators are protected, and thus that for $1<d<D-1$ we have self-correcting quantum memories, which thus exist for $D\geq 4$.

Recall that we have to define a connection scheme for excitations and a correction procedure such that the
 conditions (i)-(iv) of section \ref{sec:conditions} are satisified. The notion of connectedness of
  excitations, that is, of $(d-1)$-cells, is naturally given by the lattice. To translate this into the
   language of section \ref{sec:conditions} we just construct a graph with $(d-1)$-cells as its nodes in which nodes are connected by a link when the corresponding cells meet at a $(d-2)$-cell (notice that $d\geq 2$). Conditions (i) and (ii) are clearly satisfied. As for  condition (iii), we first notice that any connected set of excitations is closed. Some connected sets are boundaries, so that they can be switched off independently and give no further problems (i.e. they constitute $\nodes(E)$ for some Pauli operator $E$). But those which have non-trivial homology cannot be switched off independently. Instead, we have to gather several of them together to get a boundary, but only in a finite amount. Indeed, setting $\nu=h_{d-1}+1$ condition (iii) is satisfied because $h_{d-1}$ is the dimension of $H_{d-1}$ as a $\Z_2$ vector space.

Finally, we need a correction procedure, that is, the function $\corrector_X$ in \eqref{func_corr_synd_X_Z}. We define it in two steps. In the first step we divide the syndrome $\vect b$ in the components $\vect b_i$ of \eqref{decomposition_syndrome}. In the second step, for each $\vect b_i$ we find a suitable collection $c\prima$ of $d$-cells such that $\partial c\prima = c = \nodes(\vect b_i)$. The choice of the collection of $d$-cells $c\prima$ must be such that it ensures stability. In fact, we only consider those connected sets of $(d-1)$-cells that can be put inside a hypercube of length $l$ with $l\leq L$, and choose any $c\prima$ that is contained in the hypercube. This task that is always possible because $c$ has trivial homology, and it can be easily checked that it can be accomplished in polynomial time in $l$. We leave the correction of other sets of $(d-1)$-cells undefined, because it is not important.

Now that we have a correction procedure, we can check condition (iv). It suffices to show that no $Z$-syndrome $\vect b\in \Z_2^{g_2}$ that only contains  connected sets in $\nodes(\vect b)$ of size at most $L/2-1$ can be
 critical. To this end, take any $d$-cell $t$ and let us check whether $\vect b$ is $X_t$-critical. Let $s_i$ be the connected components of the set of $(d-1)$-cells in $\nodes(\vect b)\cap\partial_1 t$.
Both $t$ and these sets $s_i$ are contained in a hypercube of length $L-1$, because each $s_i$ is contained in a hypercube of length $L/2-1$. Now recall, that a syndrome $\vect b$ is criticial with respect to a
given error $E$ and encoded operator $N$, iff  the operator \eqref{trick_S} anticommutes with $N$.
Using the notation of \eqref{syndrome_topo_codes} we have that the operator of \eqref{trick_S} $\corrector_X (\vect b)X_t\corrector_X(\vect b+S_0\prima\cap\partial_1 t)$ has support in the hypercube of length $L-1$ and thus has trivial homology, hence due to \eqref{exact_closed_comm} it commutes with any
encoded $Z$-type operators $N\subset \mathcal P_Z$. This implies condition (iv) and thus the stability of $N_\stable$ for any given $N\subset \mathcal P_Z$.

\subsection{Generalized color codes}

Color codes are local codes that are constructed in a special kind of lattices called $D$-colexes that live in $D$-manifolds. We will first describe color codes without referring to homology, and then give a homological picture to fit them in the general framework given above. As in the case of toric codes, several color codes labeled with $d=0,\dots, D$ can be obtained from a given lattice, which again will display $d-1$ dimensional and $D-d-1$ dimensional excitations. However, unlike in toric codes these excitation can branch and form nets.

\subsubsection{D-colexes}\label{sec:D-colexes}

A $D$-colex $\Lambda$ is a $D$-dimensional lattice with certain colorability properties. Such lattices $\Lambda$ can be described \cite{branyons} in terms of their 1-skeleton or graph, wich is $(D+1)$-valent and has $D+1$ colorable edges. Here, instead, we choose to define a $D$-colex $\Lambda$ in terms of its dual lattice  $\Delta$. From now on $D$ is an integer, $D>0$.

\begin{defn}
A lattice $\Lambda$ is a $D$-colex if its dual lattice $\Delta$ is a simplicial lattice with $(D+1)$-colored vertices that forms a closed $D$-manifold.
\end{defn}
 Recall that an $n$-simplex is the convex hull of a set of $(n + 1)$ affinely independent points. In a simplicial $D$-dimensional lattice all the $n$-dimensional cells are homeomorphic to $n$-simplices. A closed manifold is a compact manifold without boundaries. For example, for $D=1$ it must be a circle, and for $D=2$ \cite{topologicalClifford} it could be a sphere, a projective plain, a torus, etc. That the vertices of $\Delta$ are $(D+1)$-colored means that we have attached labels to vertices from a set $Q$ with $D+1$ elements or colors, in such a way that two vertices connected by a link never share the same color. At first sight it may seem unclear whether a $D$-colex can be constructed in arbitrary manifolds, due to the colorability constraints. However, in \cite{branyons} a construction is given to obtain $D$-colexes from arbitrary lattices.
 In the case of hypercubic lattices, the construction is particularly simple.
The resulting dual lattice $\Delta$ has vertices also forming a $D$-dimensional hypercubic lattice, which we can label by their coordinates $(x_j)\in \Z^D$. The $D$-simplices are as follows.
Let $\vect e_k =: (\delta_{kj})$ be the standard basis. For each permutation
of $D$ elements $\sigma_j$, $\vect z = (z_j)\in \Z^D$ and $(d_j) \in \sset{1,-1}^D$,
there is a simplex with vertex coordinates
$$\sset{2z, 2z + d_1 \vect e_1 , 2z + d_1\vect e_1 + d_2\vect e_2 ,\dots,2z+\sum_{i=1}^D d_i \vect e_i}.$$
This lattice is $(D+1)$-colorable: the coloring $c\in \sset{0, . . . , D}$ of a
vertex $(x_i)$ is given by $c =
\sum_i(x_i \mod 2)$.

We will denote the set of all $n$-simplices in $\Delta$ as $\Delta_n$, $n=0,\dots, D$. The coloring of vertices induces a coloring of $n$-simplices. Namely, given a subset of colors $q\subset Q$ with $n+1$ elements, we define the subset $\Delta_q \subset \Delta_n$ of those $n$-simplices such that their $n+1$ vertices are labeled with the elements of $q$. A coloring for the cells of the $D$-colex $\Lambda$ is also induced. Let us write $\lambda^\ast$ for the dual $(D-n)$-simplex of a $n$-cell $\lambda$ and $\delta^\ast$ for the dual $(D-n)$-cell of a $n$-simplex $\delta$. Then if $\delta$ is a $q$-cell we say that $\delta^\ast$ is a $\bar q$ cell, with $\bar q$ the complement of $q$ in $Q$. As with simplices, the set of all $n$-cells of $\Lambda$ is $\Lambda_n$ and the set of $q$-cells is $\Lambda_q\subset \Lambda_n$, $q\subset Q$, $|q|=n$. Note in particular that vertices are colorless. Indeed, the colorability of vertices in $\Lambda$ has its own content: vertices are bicolorable --- that is, the graph of $\Lambda$ is bipartite --- if and only if the manifold is orientable. In general $\Lambda$ may be composed of $m$ connected components $\Lambda^\alpha$, $\alpha =1,\dots, m$. We will regard each of these components as a `$(D+1)$-cell', setting $\Lambda_Q:=\Lambda_{D+1}:=\sset{\Lambda^\alpha}$.

The previous notation for $\Delta$ and $\Lambda$ can be extended to individual simplices and cells. Namely, $\delta_n\subset \Delta_n$ will denote the set of $n$-simplices that are part of an $m$-simplex $\delta\in \Delta_m$, $D\geq m\geq n\geq 0$. For example, if $\delta\in \Delta_1$ is a link then $\delta_0=\sset{\epsilon,\eta}$ with $\epsilon$, $\eta$ the vertices in the ends of $\delta$. Similarly, $\lambda_n\subset \Lambda_n$ denotes the set of $n$-cells that are part of an $m$-cell $\lambda\in \Lambda_m$, $m\geq n$, and $\lambda_q\subset \Lambda_q$ denotes the set of $q$-cells that are part of a $q\prima$-cell $\lambda\in\Lambda_{q\prima}$, $q\subset q\prima\subset Q$.

An important feature of $D$-colexes is that all their $d$-cells, $0<d\leq D$, are $(d-1)$-colexes themselves. This can be easily checked in the dual simplicial lattice $\Delta$. To this end, we first choose a vertex $\delta\in \Delta_0$. If we keep only those simplices that meet at $\delta$, and reduce their dimension in one by erasing the vertex $\delta$, it is easy to check that the resulting simplicial lattice $\Delta\prima$ is dual to a $(D-1)$-colex $\Lambda\prima$ with the topology of a $(D-1)$-sphere. This is the colex that correspond to the $D$-cell $\delta^\ast$. The process can then be continued recursively to get the colexes attached to cells of smaller dimension. Thus, there is a one-to-one correspondence between subcolexes and cells. Each $d$-cell $\lambda\in \Lambda_d$ has as its boundary a $(d-1)$-subcolex. If $\lambda$ is $q$-colored, $q=\sset{q_i}\subset Q$, then its $(d-1)$-colex is composed of $q_i$-links, $i=1,\dots, d$.

Given a coloring $r$ and a cell $\lambda\in \Lambda_q$, $q\subset r\subset Q$, there exists a unique $r$-cell, denoted $\lambda^r\in \Lambda_r$, such that $\lambda\in(\lambda^r)_q$. It is natural to define the intersection of two cells $\mu\in \Lambda_q$, $\nu\in \Lambda_{r}$, $q,r\subset Q$, as the collection of disjoint $(q\cap r)$-cells $\mu\cap\nu:=\mu_{q\cap r}\cap\nu_{q\cap r}$.  When $\mu\cap\nu$ is empty or has a single element for any cells $\mu$ and $\nu$, we will say that $\Lambda$ is nice. In a nice $D$-colex, if $\mu\in \Lambda_q$ and $\lambda\in \mu_s$ with $s\subset q\subset Q$, there exists for any $r\subset Q$ with $s= q\cap r$ a unique $\nu\in \Lambda_r$ with $\mu\cap\nu=\sset{\lambda}$, namely $\nu=\lambda^r$.

\subsubsection{Color codes}

From a $D$-colex $\Lambda$ we can obtain several color codes, labeled with an integer $d$ such that $0\leq d \leq D$. As we will see, the $d$-th code is dual to the $\bar d$-th code.
The codes with $d=\bar d$ are self-dual, something that turns out to have interesting computational consequences.

To construct the $d$-th color code from a $D$-colex $\Lambda$, we first attach a qubit to each vertex $v\in\Lambda_0$. As generators of the stabilizer $\mathcal S$ we choose certain $(d+1)$-cell and $(\bar d+1)$-cell operators, as we explain next. For any $n$-cell $\lambda\in\Lambda_n$, $0\leq n\leq D+1$, we introduce the $X$ and $Z$ cell operators
\begin{equation}\label{cell_opts}
 X_\lambda := \prod_{v\in \lambda_0} X_v,\qquad Z_r := \prod_{v\in \lambda_0} Z_v.
\end{equation}
An important property of cell operators is  that for any integers $1\leq n,m\leq D+1$ with $n+m>D+1$, and for any cells $\mu\in \Lambda_n$, $\nu\in \Lambda_m$ we have \cite{branyons}
\begin{equation}\label{cell_opts_comm}
 [X_\mu,Z_\nu]=0.
\end{equation}
Also, for any $q$-cell $\lambda$, $q\subset Q$, and for any coloring $q\prima\subset q$, we have the relations \cite{branyons}
\begin{equation} \label{cell_opts_rel}
X_\lambda = \prod_{\mu\in\lambda_{q\prima}} X_\mu, \qquad Z_\lambda = \prod_{\mu\in \lambda_{q\prima}} Z_\mu.
\end{equation}

\begin{defn}
The $d$-th color code of $\Lambda$, $0\leq d\leq D$, is the stabilizer code $\mathcal C$ with stabilizer $\mathcal S$ generated by the operators of the form $X_\lambda$, $\lambda\in \Lambda_{d+1}$,  and $Z_{\lambda\prima}$, $\lambda\prima\in \Lambda_{\bar d+1}$.
\end{defn}
 From this definition it is clear that dual codes are related through the exchange of $X$ and $Z$ operators, and thus are equal up to a change of basis.

From the relations \eqref{cell_opts_rel} it follows \cite{branyons} that the number of encoded qubits is
\begin{equation}\label{topo_degeneracy}
 k=\binom D d \,h_d
\end{equation}
where $h_d=h_{\bar d}$ is the $d$-th Betti number of the manifold. The origin of this particular value will be clarified in the following sections.

Note that color codes with $d=0$ or $d=D$ are not interesting quantum codes, as they are repetition `classical' codes, which only correct either $X$ or $Z$ errors and encode a qubit per connected component of $\Lambda$. Because of this, below we will mainly focus on color codes with $0<d<D$.

\subsubsection{A simplicial homology}

In this section we will work only in the dual lattice $\Delta$,
since it is more suited to explain the underlying homological
structure of color codes. We attach to each set of
$n$-simplices $\Delta_n$, $0\leq n\leq D$, the abelian group $C_n$
of $\Z_2$ chains, together with a trivial group $C_{-1}=\sset{0}$,
and to each $n=0,\dots, D$ a boundary operator
\begin{equation}
 C_D\,\xrightarrow {\hat\partial_{n}}\,C_{n-1}
\end{equation}
defined for $n>0$ by
\begin{equation}
\hat\partial_n\, \delta := \sum_{\epsilon\in \delta_{n-1}} \epsilon,
\end{equation}
where $\delta\in\Delta_D$. The exact sequence from which the $d$-th color code is obtained is
\begin{equation}\label{exact_seq_hat}
 C_{\bar d-1} \,\xrightarrow {\hat\partial_{\bar d}^\ast}\,C_D\,\xrightarrow {\hat\partial_{d}}\,C_{d-1},
\end{equation}
where $0\leq d\leq D$. It gives rise to the homology group
\begin{equation}\label{homology_color_d}
\hat H_d := {\hat Z_d}/{\hat B_d}.
\end{equation}
It follows from \eqref{topo_degeneracy} that the homology groups $\hat H_{\bar d}$, $\hat H_d$  are isomorphic to the sum of $\binom D d h_d$ copies of $\Z_2$. In other words, we have
\begin{equation}\label{homology_color_d_vs_d}
\hat H_d \sim \binom D d H_d.
\end{equation}
The origin of the binomial coeffiecient is related to color and will be explained in the next section.

\subsubsection{Brane-nets}

We will now give a geometric interpretation to the $D$-simplex cycles discussed in the previous section in terms of branes. For any $n$-cell $\lambda\in\Lambda_n$ we define $c_\lambda \in C_D$ as
\begin{equation}
c_\lambda := \sum_{v\in\lambda_0} v^\ast.
\end{equation}
For $0<n<D$, we will now introduce boundary operators that take $n$-cells to $(n-1)$-chains in a `natural' way. For $\lambda\in \Lambda_{n}$ we set
\begin{equation}
\partial \lambda := \hat\partial_{n} \,c_\lambda.
\end{equation}
We say that these definitions are natural because $\partial \lambda$, as a set of simplices glued together, is homeomorphic to the boundary of $\lambda$ in the usual sense. Moreover, if $\lambda$ is a $q$-cell, $q\subset Q$, then $\partial \lambda$ only contains $q$-simplices. Thus, we can regard $n$-cells as pieces of $n$-branes. Indeed, coloring is important here, and we should regard a $q$-cell, $q\subset Q$, as a piece of a $q$-brane. As we will see next, such pieces can be put together to form closed $q$-branes, that is, branes without boundaries in the sense of the $\partial$ just defined, which is just a reinterpretation of $\hat\partial_n$.

Let us fix a particular color subset $q$ with $|q|=n$ and introduce the $q$-colored chain subgroup $C_D^q$, the subgroup of $C_D$ generated by all $c_\lambda$ chains with $\lambda\in \Lambda_q$. That is, the chains in $C_D^q$ represent the $q$-branes just discussed. We may then set $Z_q:=\hat Z_n\cap C_D^q$, $B_q:=\hat B_n\cap C_D^q$ and $H_q:= Z_q/B_q$ with the net result
\begin{equation}\label{homology_d_q}
H_q \sim H_n.
\end{equation}
Thus, the homology of $q$-branes for a fixed $q$ is just the usual brane homology. This result is shown in \cite{branyons} by constructing from the colex $\Lambda$ a $q$-reduced lattice in which all $n$-cells correspond to $q$-cells of $\Lambda$.

In summary, the cycles in $Z_q$ can be regarded as closed $q$-branes, which are also cycles in $\hat Z_n$. In general, cycles in $\hat Z_n$ will combine several pieces of such $q$-branes of different color, forming $n$-brane-nets \cite{branyons}. In this sense, $\hat H_n$ is a $n$-brane-net homology group, as opposed to the usual $H_n$ homology group of $n$-branes. This discussion already suggests that $\hat H_n$ contains several copies of $H_n$, but how many of them? To answer this, let us consider color subsets $q_i\subset Q$, $i=1,\dots D+2-n$, with $|q_i|=n$ and $|q_i\cap q_j|=n-1$ for $i\neq j$. It turns out \cite{branyons} that if $z_i$, $i=1,\dots D+1-n$, are $ q_i$-branes, all of them homologous in the usual sense of $H_n$ homology, then
\begin{equation}
\sum_i z_i \sim 0
\end{equation}
up to $\hat H_n$ homology, which shows that only $\binom D n$ color combinations are independent, which finally explains \eqref{homology_color_d}.

In a $d$-th color code the relevant homology groups are $\hat H_{\bar d}$ and $\hat H_d$, with the corresponding $\bar d$- and $d$-brane-nets, which are mapped to operators using \eqref{chain_opt}. The commutation rules of closed brane-net operators are purely topological.
Namely, given a closed $q$-brane-net $z\in \hat Z_q$, $|q|=d$, and a closed
$q\prima$-brane-net $z\prima\in\hat Z_{q\prima}$, $|q\prima|=\bar d$, the operators $X_z$ and $Z_{z\prima}$ anticommute if $z$ and $z\prima$ intersect and odd number of times and $q\cap q\prima =\emptyset$. Otherwise they commute.

\subsubsection{Branyon-nets}

For $d=1,\dots,D$, the Hamiltonian model \eqref{Hamiltonian_X_Z} of a $d$-th color code contains as in toric codes two kinds of brane-like excitations, $(\bar d-1)$-branyons and $(d-1)$-branyons. These are created with the brane-net operators discussed in the previous sections, in particular with open brane-nets since the excitations appear at their boundary. Then, unlike in toric codes, branyons can form nets and we must talk about branyon-nets. As branes in the previous sections, branyons can be labeled with colors. Namely, a $q$-branyon is the boundary of a $q$-brane.

We will only discuss $(d-1)$-branyons, which are related to violations of the stabilizer conditions at $(\bar d+1)$-cells. The case of $(\bar d-1)$-branyons is analogous. If $\lambda\in \Lambda_{\bar d+1}$ is a $q$-cell, we say that
the corresponding excitation is a $(Q-q)$-branyon excitation. We know that branyon excitation must be arranged in closed structures. This fact is reflected \cite{branyons} in a `conservation law' at $(\bar d+2)$ cells. Namely, for $\lambda$ a $q$-cell, $|q|=\bar d+2$, and for any $r,s \subset q$, $|r|=|s|=\bar d+1$ we have
\begin{equation}\label{conservation}
\prod_{\mu\in \lambda_{r}} Z_\mu = \prod_{\mu\in \lambda_{s}} Z_\mu
\end{equation}
which is just a special case of \eqref{cell_opts_rel}. This implies, for example, that at a given $(d+2)$ cell there cannot exist a single excited $(d+1)$-cell.

In the previous section we have seen that only $\binom D d$ color combinations are enough to label an independent set of $d$-brane operators. We will now show that we can locally transform a set of branyon excitations into a new set that only contains branyons from $\binom D d$ independent colors. By `locally' we mean that it is enough to act in the neighborhood of the undesired excitations. We start choosing a color $q_0\in Q$. We want to show that we can eliminate all $q$-branyon excitations with $q_0\in q$ or, to put it in a different way, all excitations located at cells $\lambda\in \Lambda_{\bar d+1}$ of color $q$ with $q_0\nin q$. The total set of excitations corresponds to the boundary of some $c\in C_D$, that is, they can be switched off with $X_c$. Notice that  all the $q$-cells with $q_0\nin q$ that touch a $(Q-q_0)$-subcolex are part of it. Then the restriction of $X_c$ to those $(Q-q_0)$-subcolexes that contain the undesired excitations will do the job.

Once we have reduce the branyonic excitations to the $\binom D d$ color combinations, each color may be treated independently as in the previous section. That is, the branyons of each color form boundary $(d-1)$-branes.

Finally, we note that, as in the case of generalized toric codes, one can derive \cite{branyons} from the commutation rules for closed brane operators the topological $ \pm 1$ phases appearing in branyon processes.

\subsubsection{Self-correcting models}
\label{subsubsec-self-color}
Regarding self-correction, color code models behave very much like toric code models. That is, for $d> 1$, the models with Hamiltonian \eqref{Hamiltonian_X_Z} for the $d$-th code protect $Z$-type encoded operators, and for $d< D-1$ they protect $X$-type operators. Instead of repeating the analysis already done for toric codes, we will just mention the new features that appear in color codes. In particular, we consider the correction of $X$-type errors.

The first difference is the lattice, but any periodic $D$-colex will do, and the existence of periodic lattices follows from the procedure given in \cite{branyons} to construct $D$-colexes from arbitrary lattices. The second difference is that now we deal with branyon-nets instead of branyons, but this does not make such a big difference due to the discussion in the previous sections. In this regard, we first need a notion of connectedness of $(d+1)$-cells. This is already implicit, for example, in \eqref{conservation}: we construct a graph with $(\bar d+1)$-cells as its nodes in which nodes are connected by a link when the corresponding cells are part of the same $(\bar d+2)$-cell (notice that $d\geq 2$). As for the $\corrector$ function, we transform each connected set of excitations using the procedure of the previous section into a collection of $(d-1)$ boundaries of $\binom D d$ colors, so that each color can be independently treated exactly as in a toric code.

\subsection{The transversality properties of color codes}

Color codes can have very special transversality properties when the lattice and its topology are suitably tuned. In this section we will treat this issue in detail. To this end, we start explaining a simple way to consider $D$-colexes in manifolds with boundary. We then describe a particular topology for the manifold: in the shape of a D-simplex in which each face has different properties. Finally, we analyze the transversality properties of the resulting simplicial codes.

\subsubsection{Boundaries}

Up to now we have only considered colexes $\Lambda$ in closed $D$-manifolds. However, richer color codes can be obtained by introducing suitable boundaries in the manifold \cite{topologicalClifford, withoutBraiding}. The easiest way to describe such boundaries is as collections of missing cells in the lattice. For a missing cell in $\Lambda$ we really mean a missing generator of the stabilizer. When the removed generator is independent from the remaining ones, the number of encoded qubits $k$ gets increased \eqref{encoded_qubits}. This can be understood also in homological terms. Consider a $q$-brane $c\in C_D$ with a boundary, $|q|=d$. Given an encoded state $\ket \psi$, the state $X_c\ket \psi$ will have stabilizer violations at those $(\bar d+1)$-cells that form the boundary of $c$. But if we eliminate the generators that correspond to these $(\bar d+1)$-cells then $X_c$ will belong to the stabilizer and thus we should consider $c$ a closed $q$-brane. This way, as cells are eliminated we are introducing in the manifold areas in which suitable branes can have a boundary and still be closed. It is as if we had removed a piece of manifold in these aereas, thus creating a boundary. Such an introduction of a boundary does not change homology only in the way just discussed. In addition, when we remove a $q$-cell the corresponding generator, which can be regarded as a $d$-brane, transforms into an encoded operator in $\mathcal N - \mathcal S$. Thus, this $d$-brane should no longer be considered a boundary, but rather a nontrivial cycle.

Although for our purposes it will be much more convenient to describe boundaries as big missing cells in the lattice, this would not be convenient at all in an implementation of the model. Fortunately it is possible to adopt a different approach, as discussed in \cite{condensate_borders} in the context of 2D toric code models and also in \cite{condensation_confinement} in a more general framework. In this alternative approach boundaries appear at interfaces between the topologically ordered phase and other phases in which some of the branyons are condensated.

\subsubsection{Simplicial color codes}

There exists a very special class of color codes which live in simplex-shaped lattices. For $D=2$ we have triangular codes \cite{topologicalClifford}, which bear this name because their lattice is a triangle: each side of the triangle corresponds to a different kind of boundary. For $D=3$ we have tetrahedral codes \cite{withoutBraiding}, and more generally for arbitrary $D$ we may consider simplicial codes. Instead of giving a detailed description of the properties of each of the boundaries in a simplicial code, it is just easier to give a constructive procedure to obtain such codes, as we do next.

Suppose that we have at our disposal a $D$-colex $\Lambda$ with the topology of a $D$-sphere, $D\geq 2$. Since spheres are homologically trivial for $D\geq 2$, $\Lambda$ yields trivial color codes with $k=0$, no encoded qubits. Let us now choose a vertex of $\Lambda$ and proceed to remove the corresponding physical qubit and all the $(d+1)$- and $(\bar d+1)$-cells that meet at this vertex. The resulting lattice looks like a $D$-simplex, because it is a $D$-ball with its boundary $(D-1)$-sphere divided in the $D+1$ regions that where in contact with each of the $D+1$ removed $D$-cells.  It can be checked using \eqref{cell_opts_rel} that only two independent generators of $\mathcal S$ are eliminated in the process, one of $X$ type and the other of $Z$ type. Thus \eqref{encoded_qubits}, all color codes derived from the new lattice $\Lambda\prima$ encode exactly one qubit. Since the original lattice has an even number of qubits \cite{withoutBraiding}, it follows that the operators
\begin{equation}\label{transv_X_Z}
 \bar X := X^{\otimes n}, \qquad \bar Z := Z^{\otimes n},
\end{equation}
with $n$ the total number of remaining qubits, are the encoded $X$ and $Z$ operators: they belong to the normalizer and
\begin{equation}
 \sset{\bar X, \bar Z}=0.
\end{equation}

It is easy to construct the smallest example of a simplicial color code for each dimension $D\geq 2$. We start with a $(D+1)$-hypercube, in which we color parallel links with the same color to get \cite{branyons} a $D$-colex with the topology of a $D$-sphere. Then as explained above we eliminate one of its vertices and all neighboring cells. As a result we get a color code with $2^{D+1}-1$ physical qubits in which all $n$-cells have $2^n$ vertices.

\subsubsection{Transversal gates}\label{sec:transversal gates}

We will now consider the tranversality properties of color codes. As it turns out, simplicial colexes give the most suitable color codes for the transversal implementation of quantum gates. Thus, we restrict our discussion to them.

First, it is clear that the transversal implementation of the logical $X$ and $Z$ operators is always possible, since they take the form \eqref{transv_X_Z}. The same reason makes transversal measurements of $X$ and $Z$ are always possible. Similarly, the two-qubit controlled not (CNot) gate
\begin{equation}
C_X := \ketbra 0 \otimes I + \ketbra 1 \otimes X
\end{equation}
can always be implemented transversally on pairs of equal simplicial codes by applying it to each pair of equivalent qubits. The one-qubit Hadamard gate
\begin{equation}
H := \frac 1{\sqrt 2} \begin{bmatrix} 1 & 1\\ 1 & -1\end{bmatrix}
\end{equation}
is more restrictive. As it exchanges $X$ and $Z$ operators, it can only be applied to dual codes, those with $\bar d=d$. Again, it can be obtained by applying a Hadamard to each of the physical qubits independently.

We want to consider in general the family of one-qubit phase gates
\begin{equation}
R_k:= \ketbra 0 + \exp(\frac {\pi i} {2^k}) \ketbra 1,
\end{equation}
where $k$ is a non-negative integer. The case $k=0$ is just the $X$
gate. The case $k=1$ is important since the gates $\sset{C_X, H,
R_1}$ generate the Clifford group. Still more interesting is the
case $k=2$, because $C_X$ and $R_2$, together with $X$ and $Z$
intializations and measurements, are enough for universal quantum
computation\cite{universal_root_i}. To study in detail when these
gates can be implemented transversally on simplicial color codes, we
will introduce next the notion of $(j,k)$-good colexes.

From now on, $j, k$ are integers with $0\leq j\leq D$, $k\geq 0$.
\begin{defn} \label{defn_goodness}
A $D$-colex $\Lambda$ is $(j,k)$-good if
\begin{equation}
\paratodo c\in\hat B_j,\qquad |c|\equiv 0 \mod 2^{k+1}.
\end{equation}
\end{defn}
Note that all colexes are $(j,0)$-good. From the first expression for encoded states in \eqref{code_states} and the fact that simplicial colexes have an odd number of physical qubits it follows that simplicial color codes obtained from $(d,k)$-good colexes allow the transversal implementation of $R_k$ gates. More exactly, if we apply $R_k$ to each physical qubit independently we will get a logical $R_k^s$ gate, with an odd power $s$. But since $s$ and $2^{k+2}$ are coprime, there exists an integer $r$ such that $R_k^{sr}=R_k^{2^{k+2}+1}=R_k$.

The problem with definition \ref{defn_goodness} is that it is not really operative. But we can give a simpler characterization of $(j,k)$-goodness. For its proof, see appendix \ref{appendix_goodness}

\begin{thm}\label{thm_goodness}
A nice $D$-colex $\Lambda$ is $(j,k)$-good for some $\,\,\,\,\,k>0$ if and only if $j\prima:=2j-D\geq 0$, all $j$-subcolexes are $(j\prima,k-1)$-good and
 \begin{equation}
\paratodo \lambda\in\Lambda_{j+1},\qquad |\lambda_0|\equiv 0 \mod 2^{k+1},
\end{equation}
Moreover, the $\Leftarrow$ implication holds even if $\Lambda$ is not nice.
\end{thm}

Since $(0,k)$-good colexes only exist for $k=0$, by repeatedly applying theorem \ref{thm_goodness} it follows that a $(j,k)$-good and nice $D$-colex may exist only if
\begin{equation}
 j\geq \frac k {k+1} D.
\end{equation}
Several conclusions may be derived from this result. For example, that dual codes only allow the transversal implementation of the $R_1$ gate. It could not be other way else, since if a dual code allowed the transversal implementation of $R_2$ we would get a stabilizer code with a universal set of transversal unitary gates, something impossible \cite{no_great_codes}. For fixed $\bar d$, there is a minimal dimension $D$ for which $(d, k)$-good nice $D$-colexes exist, namely
\begin{equation}\label{dimension_for_k_gate}
D\geq (k+1) \bar d
\end{equation}
In particular, setting $\bar d=1$ it follows that the minimal dimension $D$ for which $R_k$ may be transversally implementable in a color code is $D=k+1$.

Once the bound \eqref{dimension_for_k_gate} is established, it only rests to show that it is saturated by some lattice. But the simplest examples of simplicial lattices given in the previous section do indeed saturate the bound. The next question is whether periodic lattices that saturate it exist. This is indeed the case for $D=2,3$. 
 
As for higher dimensions, we are particularly interested in the case $D=6, k=2, \bar d=2$. Indeed, we need $\bar d>1$ if the system is to be self-correcting, and $k>1$ to get universal computation with transversal operations. Thus, $D=6$ is the lowest dimensionality that puts the two features together. As it can be readily checked, the construction of section \ref{sec:D-colexes} provides a suitable periodic 6-colex.

\section{Self-correcting quantum computers}
\label{sec:self-qc}

\subsection{Initialization from thermal states}
\label{subsec:init}

\subsubsection{Initialization with a external field}

In the previous sections we have studied how quantum memories resilient to thermal noise can be obtained from local quantum models in 4D and above, both from toric codes and color codes. However, a quantum memory is of no much use if we cannot initialize it. We will now sketch a possible procedure to initialize the self-correcting topological quantum memories as long as we can control the couplings of the local quantum Hamiltonian --- without any sort of physical qubit addressing.In particular, we will consider toric and color code models with $d>1$, so that the dressed version of $Z$-type encoded operators $N$ are protected, and give a procedure to initialize all such operators to $N=1$ with a probability of failure exponentially small in the system size. For $D=d=2$, this is just the usual initialization of a 2D ferromagnet with an external magnetic field.

Our starting point is a modification of the Hamiltonian \eqref{Hamiltonian_X_Z}. We introduce single qubit Zeeman terms and some couplings to reflect the fact that now the strength of the different terms can be controlled:
\begin{equation}\label{Hamiltonian_X_Z_zeeman}
H=-g_X\sum_{i=1}^{l} X_{p_i}-g_Z\sum_{i=1}^{m} Z_{q_i}-g_0 \sum_{s} Z_s,
\end{equation}
where $s$ runs over all qubits and $g_X, g_Z, g_0\geq 0$. Before considering the details, let us sketch the initialization procedure. We start from a thermalized state with $g_Z=g_0 = 1$ and $g_X=0$. Then we switch off the Zeeman terms. Finally, we `carefully' switch on the $X$-type stabilizers till $g_1=1$, as discussed below. Notice that the Hamiltonian is exactly solvable at all times, since all the terms with non-zero coupling at a given time commute. Moreover, all the terms commute with $Z$-type encoded operators, precisely the ones that we want to initialize with eigenvalue $1$.

We first want to show that, for temperatures $1/\beta$ below a critical value, the initial thermalized state satisfies, for any encoded operator $N\in \mathcal N \cap \mathcal P_Z$ and some $t>0$,
\begin{equation}
1-\langle N_\stable \rangle_\beta \leq e^{-\delta L}.
\end{equation}
Recall that $g_X=0$ and $g_Z=g_0=1$. Both for toric and color codes we already have a graph of the excitations that correspond to $g_Z$ terms, where nodes are labeled by the $q_i$ elements of the lattice. We enlarge this graph by adding a node per physical qubit, to represent the excitations that correspond to the Zeeman $g_0$ terms. We also add a link from the node of the physical qubit $s$ to the node of $q_i$ if $\sset{X_s,Z_{q_i}}=0$, and a link between the nodes of two physical qubits $s$ and $s\prima$ if they are both linked to some $q_i$. For example, in a toric
self-correcting code with $d=4$ the $q_i$  nodes correspond to edges and the $s$ nodes to plaquettes, so that the links of the graph just represent the usual notion of connectedness for plaquettes and edges in a lattice. The Zeeman  term is responsible for the tension of membranes built of  plaquettes, while the Kitaev term for the tension of the loops, i.e., boundaries of the membranes.

It is then clear that any connected set of excitations can be independently switched off with a suitable operator in $\mathcal P_X$. We can then repeat the reasoning of section \ref{sec:conditions} to show that those excitation configurations which contain connected sets of size bigger than, say, $L/2$ are exponentially suppressed with the size of $L$. Therefore, it suffices to observe that the rest of configurations belong to the sector with $N_\stable=1$ for any $N\in \mathcal N\cap \mathcal P_Z$, as we detail next using the language of section \ref{section_homology_codes}. First, the eigenstates of the Hamiltonian may be written as $\ket a_0$ for some $a\in A_1$. The corresponding set of excitations corresponds both to the chain $a$ and its boundary $\partial_1 a\in A_0$. But $\corrector(S_2\prima\cap \partial a)=X_{a\prima}$ for some $a\prima\in A_1$ such that both $a$ and $a\prima$ lie in the same hypercube of length $L/2$, and thus $X_aX_{a\prima}\in \mathcal S$.  The result follows.

Once we have shown that the initial thermal state is in the desired sector, it is enough to check that this continues to be true as we switch off and on the interactions as indicated. This follows from the fact that $g_Z=1$ at all times, as long as we keep the temperature of the system stable. In this regard, the process of switching on $g_X$ could be a problem since at start we have arbitrarily many violations of the $X_{p_i}$ stabilizers and if $g_X$ raises too fast we will put a lot of energy in the system. To avoid this, we require that the switching on is done slow enough for the system to stay close to the thermal state. In this regard, we assume that the thermalization time is independent of the system size, so that the time required to perform a gate does not scale either. This is a natural assumption for the systems considered here given the nature of their excitations, at least as long as the coupling with the environment is uniform.

\subsubsection{Initialization by code deformation}\label{sec:code_deformation}

An alternative way to initialize a topological quantum code is to use the code deformation approach as discussed in \cite{deformation1,deformation2}. Without going into any details, the idea is to start from a code with no encoded qubits, and to change its geometry progressively to introduce an encoded qubit and at the same time initialize it in a well-defined manner. Notice that the geometry is dictated by the stabilizer generators, and this in turn are the terms in the Hamiltonian, so that we are required to control the couplings of the protecting Hamiltonian at least with a certain degree of spatial resolution, even if no detailed control is required.

\begin{figure}
  \centering
  \includegraphics[width=8cm]{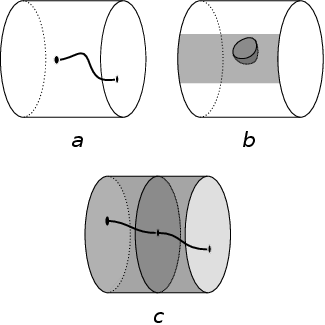}
  \caption{Initialization of a 3D toric code by changing the boundary conditions over time. a) At start, open string operators with their endpoints on any boundary surface (as the one displayed) are stabilizers. There are no encoded qubits. b) Boundary conditions change over a region (shaded) that extends as time passes. On this region string operators cannot end. Instead, membrane operators with their boundary in the region (as the one displayed) are stabilizers. c) When the region with the alternate boundary conditions becomes non-simply connected, an encoded quit appears. The $X$ logical operator is the membrane shown, whereas the $Z$ logical operator corresponds to the string. The latter is initialized to 1 by continuity, as such a string operator was before a stabilizer.}
\label{fig:deformation}
\end{figure}

If the final encoded quit should be initialized to $Z=1$, the logical operator $Z$ should correspond (up to the deforming process) to a stabilizer in the original trivial code. Also, along the whole process the geometry must be such that the $X$ logical operator, when it exists, has a global support, ensuring that the encoded quit is topologically protected from bit-flip errors. On the other hand, it will not be protected at all times from phase-flip errors, but just as in the initialization with Zeeman terms such errors are immaterial.
As an example, figure \ref{fig:deformation} shows how this could be done in a 3D toric code by changing boundary conditions (stabilizers) over time.

There is thus a parallelism in these two ways to initialize the quantum memory, in the sense that they rely on ``unprotecting'' at some point the logical operator that provides the initialization basis. It is unclear whether this is a general feature in the initialization of a self-correcting (or maybe just self-protecting) quantum memory or a particularity of our approach.

\subsection{Measurements}

The fact that we are dealing with CSS-like codes implies that we may perform transversal $X$ and $Z$ measurements. Typically, measuring a physical qubit in the $Z$ basis requires an interaction $Z\otimes H_{app}$ with a mesuring apparatus. In this regard, during the measurement we can switch off the $X$-type stabilizers, which do not commute with the measuring Hamiltonian. Switching off this part of the protecting Hamiltonian is not a source of problems since the corresponding errors do not affect the measurement.

Alternatively, code deformation could also be used to perform the measurements.

\subsection{Transversal gates}

In section \ref{sec:transversal gates} we showed that 6D color codes are enough to have a set of universal transversal operations. Local CNots require an additional dimension to remain local, so that we need a total of seven spatial dimensions.

In order to perform transversal gates we have to temporarily modify the Hamiltonian of the system. This means adding new terms that will not commute with the protecting ones and thus make the analysis difficult. In any event, when a gate is half-way of being completed the system will in general be in a state with a high energy with respect to the protecting Hamiltonian. This could easily be a source of problems. To circumvent them, we propose to apply gates, that are fast in comparison with 
the protective Hamiltonian.  Thus, during application of gates, the protective Hamiltonian can be neglected.

Notice that we can represent the faulty gate as an ideal gate preceded by some noise operation, so that it is enough to consider a situation with a trivial gate (identity), i.e., qubits are subject solely to the interaction with the environment. Thus, the process can be seen as a noisy channel, so that we can use the results of section \ref{sec:noise_resilience}, which tell us that the quantum memory can resist a small amount of noise. Indeed, the results in that section where obtained for a depolarizing channel acting independently on each qubit, but they can be generalized to more complicated noises as long as correlations are local. 

Let us note, that since very fast gates may result in a large noise \cite{AlickiLZ} 
to make separation between time scales of the gate and the protective Hamiltonian,
we have to tune down the coupling constant of the latter one, rather than make gate faster.
Then however, the ratio $T/\Delta$ (where $\Delta$ is minimal separation between the energy levels
of protective Hamiltonian) will become larger, and the threshold 
for resilience against white noise will decrease. Therefore in order to keep the noise produced 
by gate below that threshold, we shall tune down temperature. 
In this way, if we are given  good enough gates, we are able to choose 
coupling constant of protective Hamiltonian and the temperature in such a way, 
that the gate is applied to logical qubit with arbitrary good fidelity. 

However, this is not really the end of the story. Even if we could perform transversal gates perfectly and in an instant, the gates themselves will increase the number of excitations. In the case of single-qubit gates we know that excitations cannot spread much, so this is not a big problem. The case of CNot gates is worse, because they copy errors (and thus excitations) from one code to another. It is intuitively clear, however, that as long as excitations are scarce enough this will not affect the encoded qubits. A rigorous approach to this problem requires using similar arguments to those in sections \ref{sec:conditions} and \ref{sec:noise_resilience}, but considering two copies of the $\Gamma_L$ lattices together.

In summary, the application of a transversal gate will put some amount of energy per volume unit in the system. As long as the temperature is low enough and the energy not too much, this will not produce an error. Of course, we have to wait some time between gates, so that the injected energy can be dissipated.

\section{Conclusions}

We have provided a quantum-topology based solution to the problem
of constructing a self-correcting QC. One of the main problems we have
to face is the existence of tensionless string excitations (errors) 
that propagate throughout the quantum device causing the QC to collapse.
Thus, a mechanism for reabsortion of errors is necessary and a natural
way to achieve this in a topological scenario is by increasing
the spatial dimension, but this has to be done with 
an appropriate  class of topological codes like TCCs which 
provide a complete set of tools for a self-correcting QC.
This situation is reminiscent  of another fundamental problem in gauge theories,
i.e., color confinement in QCD. In fact, there exists the string picture 
of confinement based on string tension on quarks \cite{confinement}. We may conjecture
that a possible solution to the critical dimension of the self-correcting
QC is to find appropriate confinement mechanisms, i.e., we arrive at the
following relationship:
\begin{equation*}
\text{Self-Correcting QC} \rightleftharpoons \text{Confinement Problem}.
\end{equation*}
The solution of this confinement problem is an open question in topological quantum computation.

Related to this central problem, we have constructed a model for a self-correcting quantum computer
based on topological stabilizer codes. The model, to be local,
requires $7$ spatial dimensions ---the codes require 6 dimensions and an extra dimension is needed to keep CNOT gates local.
Our model consists of 1) a lattice of qubits exhibiting proper topological
and geometrical properties so that 2) a code on this lattice
allows a universal set of transversal gates (including measurements) and
3) a protecting Hamiltonian, with controllable coupling constants.
We have  described the procedures of initialization and measurement of the quantum computer, and how gates should be performed. For the latter, 
we have also proposed an initialization method based on code deformations that avoids switching on/off Hamiltonians \cite{comment_switching}.

We have then obtained two rigorous results concerning the self-protection of such a quantum computer.  First, we have shown within the weak coupling limit regime
that the memory is stable, generalizing thereby
results of \cite{AlickiHHH-Kitaev}. Second, we have
shown that if the memory is exposed to a local noise for a short time the quantum information is preserved.
Our further results are physically well motivated and established.
These comprise the initialization scheme and a reasoning showing
that we can perform gates on protected qubits with high fidelity.

We would like to emphasize that as the formulation of a self-correcting
quantum computer is a challenge, we decided to relax the number of spatial
dimensions D in which our proposals are formulated. The point is to
address this problem as a fundamental issue regardless of the
dimensionality. Once we have seen that this is possible under the
circumstances considered in this paper, then we may reflect on possible
ways to lower the dimensionality D where the lattices are embedded. To
this end, we should consider our models as simplified models as far as the
types of degrees of freedom is concerned. We have only used spin 1/2 or
qubits. It is an open question to see whether by including additional
degrees of freedom like fermions or bosons, we could induce interactions
that may lower the dimension in which self-correcting quantum computers
are possible \cite{Tcodes_2}.
A similar possibility is to consider models with generalized spins and
discrete non-abelian gauge symmetry groups  \cite{BombinMD08-nonabelian} which are 
found to posses string tension and confining properties. Generally speaking, 
there are instances of quantum theories whose critical
dimensions for being stable can be lowered by introducing additional types
of degrees of freedom or a combination thereof. This could also be the
case here and this possibility we leave it open, but with the new insight
gained here, we see that at least with simpler models it is possible to
achieve the property of self-correctedness.

Finally let us note that our results are in  contradiction with Ref. \cite{Alicki09-perpetuum},
where it is argued that  a quantum computer could be used to draw work from a heat bath.
It would be worthwile to get more insight into this problem, especially regarding the 
points which are more physically stablished in our approach, particularly the limitations in the access to the quantum memory.

\section{Acknowledgements}
We are grateful to Robert Alicki, for asking questions leading to this paper, and numerous discussions. M.H. thanks Jonathan Oppenheim for discussions. H.B. and M.A.M.-D. acknowledge
financial support from a PFI grant of EJ-GV, DGS grants under contracts, FIS2006-04885, and the ESF INSTANS 2005-10.  R.W.C. and M.H. are supported by EC IP SCALA and by Polish Ministry of Science and Higher Education through Grant  No. NN20223193. R.W.C. also acknowledges support from the Foundation of Polish Science (FNP). The support of Polish research network LFPPI is also acknowledged. Part of this work was done in National Quantum Information Centre of Gdansk. Part of this work was initiated during the Madrid 2008 conference on
the Mathematical Foundations of Quantum Control and Quantum Information Theory (Fundacion Areces).

\appendix
\section{Relation between fidelity and autocorrelation function: proof of proposition \ref{prop:fid_virt}} \label{appendix_fidelity}

Denote $X=P_+-P_-$, with $P_+ + P_-=I$. From the definition of 
$F(\psi;\Lambda^*_{\virt,anc},\rho_{anc})$ we get 
\be
F(\psi_+;\Lambda^*_{\virt,anc},\rho_{anc})=\tr \bigl(I_\virt\ot \rho_{anc} P_+\Lambda(P_+)\bigr)
\label{eq:fid_P_plus}
\ee
and similarly  for  $F(\psi_-;\Lambda^*_{\virt,anc},\rho_{anc})$.
On the other hand, using that $\Lambda$ preserves the identity, 
we get  
\be
 X \Lambda(X)= 2(P_+ \Lambda(P_+)+P_- \Lambda(P_-))-I
\ee
This, together with Eq. \eqref{eq:fid_P_plus} implies the required formula.

\section{A bound for connected collections}\label{appendix_connected}

We want to show that, in a graph with at most $\mu$ links meeting at each node, the number of connected collections of nodes of size $k$ that contain a give node $n$ is bounded by $e^{\tau k}$ for a suitable $\tau$ that only depends on $\mu$. The idea of the proof is to express a collection as a word $W$ in a suitable alphabet with $\mu+1$ letters. For a given node, each neighboring node is identified by one of the first $\mu$ letters, and there is an extra ``back" letter. A valid word can have at most $l/2$ ``back" letters among its first $l$ letters, and we construct a collection from a valid word $W$ as follows. At each step we keep an ordered list of nodes $(n_1,\dots, n_I)$. At start the list is $(n)$, so that it has a single element. At the $j$-th step, if the $j$-th letter of $W$ is not a ``back" letter, we add the corresponding neighboring node of $n_I$, the last element of the list. If instead it is a ``back" letter, we remove the last node in the list. The collection is formed by those nodes that were in the list at some point in the process, so that clearly at most $2k-3$ letters are necessary to encode a collection of size $k$ and thus we can take $\tau=2\log (\mu+1)$.

\section{Proof of theorem \ref{thm_goodness}}\label{appendix_goodness}

First we need some extra notions for those cases in which $j\prima\geq 0$. Recall that each $\lambda\in\Lambda_{j+1}$ is a $j$-colex $\Lambda\prima$. We introduce a group homomorphism $\funcion {h_\lambda} {C_D}{C\prima_{j}}$ that restricts $c\in C_D$ to $C\prima_{j}$, the $j$-simplex chain group of $\lambda$. In particular for $c=\sum_{v\in V}{v^\ast}$ with $V\subset \Lambda_0$ we set $h_\lambda(c):=\sum_{v\in V\prima} v^\ast$ with $V\prima = V\cap \lambda_0$. This homomorphism maps $\hat B_j$ to $\hat B\prima_{j\prima}$. To check this, it suffices to show that for any $\mu\in \Lambda_{j+1}$ we have $h_\lambda(c_\mu)=\sum_{\nu\in N} c_\nu$ for some $N\subset \lambda_{j\prima+1}$. But if $\mu\cap\lambda=\sset{\nu^i}$ we just set $N=\bigcup_i (\nu^i)_{j\prima+1}$.

When $\Lambda$ is nice, we can in addition define a group homomorphism $\funcion {h^\lambda} {\hat B\prima_{j\prima}}{\hat B_j}$ such that for every $c\prima \in \hat B\prima_{j\prima}$ we have $h_\lambda(h^\lambda(c\prima))=c\prima$. For each $\nu\in\Lambda_{j\prima}$ we set $h^\lambda(c\prima_\nu) :=c_{\nu^q}$, where $q=Q-(r-s)$ with $s,r\subset Q$ such that $\nu\in \Lambda_s$, $\lambda\in\Lambda_r$. Here $c\prima_\nu$ is meant to denote $h_\lambda(c_\nu)$.

We can now prove each of the implication directions:

\noindent $\Rightarrow$ / First, for any $\lambda\in\Lambda_{j+1}$ we have $c_\lambda\in\hat B_j$, so that $|\lambda_0|=|c_\lambda|\equiv 0 \mod 2^{k+1}$. Suppose that $j\prima\leq 0$. Then for any two colorings $q,r\subset Q$ with $|q|=|r|=j+1$, $|q\cap r|=\emptyset$ and for any $v\in\Lambda_0$ we get, using the niceness of $\Lambda$, that $|c_{v^q}+c_{v^r}|=|c_{v^q}|+|c_{v^r}|-2\equiv -2 \mod 2^{k+2}$, a contradiction. Finally, suppose that $\lambda\in\Lambda_{j+1}$ is not $(j\prima, k-1)$-good as a $j$-subcolex. Then, with the same notation as above, there exists a chain $c\prima\in \hat B\prima_{j\prima}$ with $|c\prima|\not\equiv 0 \mod 2^{k}$. For $c=h^\lambda(c\prima)$ we have by assumption $|c|\equiv 0 \mod 2^{k+1}$, but then $|c+c_\lambda|=|c|+|c_\lambda|-2|c\prima|\not\equiv 0 \mod 2^{k+1}$. Since $c,c_\lambda\in \hat B_j$, this is a contradiction.

\noindent $\Leftarrow$ / It suffices to show that for any $c\in \hat B_j$ with $|c|\equiv 2 \mod 2^{k+1}$ and for any $\lambda\in \Lambda_{j+1}$ we have $|c+c_\lambda|\equiv 0 \mod 2^k$. But for $c\prima=h_\lambda(c)$ we get $|c+c_\lambda|=|c|+|c_\lambda|-2|c\prima|$ and since $|c\prima|\equiv 0\mod 2^k$ the result follows.




\begin{references}
\bibitem{comment_switching} 
 Let us note in passing that we have also proposed to temporarily switch off the protecting Hamiltonian so that we can avoid problems due to the combined effect of the protecting Hamiltonian and the one needed to perform the gates.
\end{references}
\end{document}